%% file: main.tex
\documentclass[journal=jctc,manuscript=article]{achemso}

\usepackage[version=3]{mhchem} 
\usepackage{subfiles}
\usepackage{amsmath}
\usepackage{physics}
\usepackage{graphicx}
\usepackage{dcolumn}
\usepackage{bm}
\usepackage{booktabs}

\author{Kyle Bystrom}
\email{kylebystrom@g.harvard.edu}
\author{Boris Kozinsky}
\email{bkoz@seas.harvard.edu}
\affiliation{Harvard John A. Paulson School of Engineering and Applied Sciences}

\title{CIDER: An Expressive, Nonlocal Feature Set for Machine Learning Density Functionals with Exact Constraints}

\abbreviations{DFT,ML,KS,HF,XC}
\keywords{Density Functional Theory, Machine Learning}
\SectionNumbersOn

\newcommand*\diff{\mathop{}\!\mathrm{d}}

\begin{document}
\begin{singlespace} 
\begin{tocentry}

\includegraphics{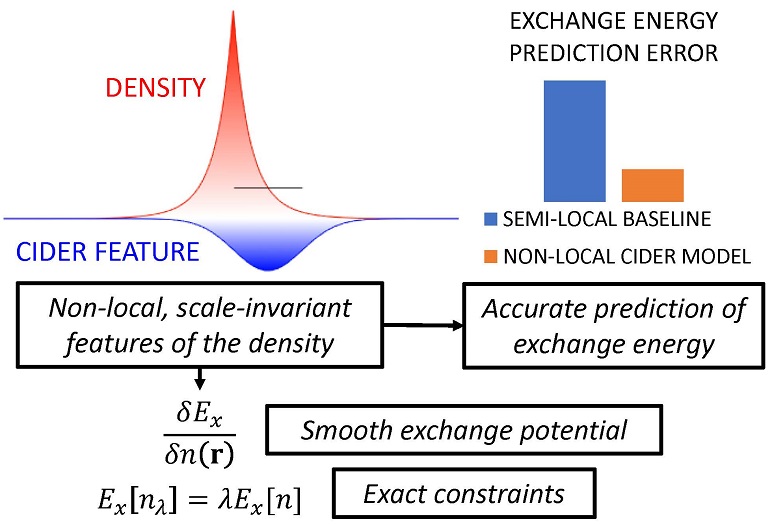}

\end{tocentry}

\begin{abstract}
  Machine learning (ML) has recently gained attention as a means to develop more accurate exchange-correlation (XC) functionals for density functional theory, but functionals developed thus far need to be improved on several metrics, including accuracy, numerical stability, and transferability across chemical space. In this work, we introduce a set of nonlocal features of the density called the CIDER formalism, which we use to train a Gaussian process model for the exchange energy that obeys the critical uniform scaling rule for exchange. The resulting CIDER exchange functional is significantly more accurate than any semi-local functional tested here, and it has good transferability across main-group molecules. This work therefore serves as an initial step toward more accurate exchange functionals, and it also introduces useful techniques for developing robust, physics-informed XC models via ML.
\end{abstract}

\subfile{body/introduction}

\subfile{body/cider}

\subfile{body/gps}

\subfile{body/methods}

\subfile{body/results}

\subfile{body/conclusion}

\appendix

\subfile{body/appendix_extra_math}

\subfile{body/appendix_extra_math2}

\subfile{body/appendix_mapping}

\subfile{body/appendix_derivs}

\begin{acknowledgement}

The authors thank Prof. Donald Truhlar for helpful discussions regarding use of the Minnesota 2015 Database. This work was supported by the US Department of Defense MURI under Award No. N00014-20-1-2418 and by the STC Center for Integrated Quantum Materials, NSF Grant No. DMR-1231319. KB was supported by the Department of Defense (DoD) through the National Defense Science \& Engineering Graduate (NDSEG) Fellowship Program.

\end{acknowledgement}

\begin{suppinfo}

Training and validation details for the CIDER functionals (S1), additional benchmarking details for the Gaussian process models (S2), a linear-scaling algorithm for CIDER in Gaussian-type orbital basis sets (S3), discussion of different possible choices for the exchange energy density (S4), and a discussion of the difference between KS and HF exchange energies (S5).

\end{suppinfo}

\bibliography{references}

\end{singlespace} 
\end{document}


\maketitle

\beginsupplement

\section{Validation of CIDER Functional and Hyperparameter Selection\label{si:cider_val}}

\subsection{Selection of Parameters for CIDER Functional}

The normalization of the features for CIDER (eq 19) requires a choice of hyperparameters to be made.  $\gamma_x$ was set to $0.016 (6\pi^2)^{2/3}$, a value derived by fitting to Hartree-Fock exchange energies~\cite{Becke1986} and used in B97-type functionals~\cite{Mardirossian2014,Mardirossian2016}. $c_3$, $c_9$, and $c_{10}$ were set to $1/2$, and $\gamma_{0a,b,c}$ were all set to make the descriptor vanish for the HEG. Therefore, for Version A, $\gamma_{0a,b,c}=1/2$, while for Version C $\gamma_{0a,c}=1/2$ and $\gamma_{0b}=1/3$. $\gamma_1$ and $\gamma_2$ were set heuristically based on experiments with optimizing parameters in \textsc{GPyTorch}~\cite{Gardner2018} and by observing the range of values $G_{01}$ and $G_{02}$ took over the noble gas atoms. The following values were used:
\begin{itemize}
    \item WIDE: $\gamma_1=0.025,\gamma_2=0.015$
    \item MEDIUM-WIDE: $\gamma_1=0.02,\gamma_2=0.0075$
    \item MEDIUM-TIGHT: $\gamma_1=0.05,\gamma_2=0.02$
    \item TIGHT: $\gamma_1=0.30,\gamma_2=0.15$
\end{itemize}

\subsection{Model Selection}

\begin{table}
    \begin{center}
    \caption{RMSE of atomization exchange energy (AEX) and total exchange energy (EX) for different functionals in kcal/mol.}
    \label{tab:cider_val}
\begin{tabular}{lrrrrrr}
\toprule
{} &  AEX Closed &  EX Closed &  AEX Open &  EX Open &  AEX Total &  EX Total \\
Functional                        &             &            &           &          &            &           \\
\midrule
LDA (Slater)$^a$                  &         167 &       1872 &        79 &     1529 &        146 &      1773 \\
B88~\cite{Becke1988Functional}    &          64 &         90 &        44 &       53 &         59 &        80 \\
BR89~\cite{Becke1989,Proynov2008} &          93 &        398 &        53 &      310 &         83 &       373 \\
CHACHIYO~\cite{Chachiyo2020}      &          56 &         85 &        41 &       47 &         52 &        76 \\
MBEEF~\cite{Wellendorff2014}      &          53 &        704 &        37 &      561 &         49 &       663 \\
MN15-L~\cite{Yu2016MN15L}         &          47 &         73 &        17 &       70 &         40 &        72 \\
PBE~\cite{Perdew1996}             &          75 &         81 &        48 &       71 &         68 &        78 \\
SCAN~\cite{Sun2015SCAN}           &          62 &        119 &        40 &       88 &         56 &       111 \\
TM~\cite{Tao2016}                 &          55 &        136 &        41 &       68 &         51 &       119 \\
\hline
GP-SL$^b$                         &          94 &        277 &        52 &      157 &         84 &       246 \\
\hline
GPXA-W$^c$                        &          14 &         13 &        13 &       11 &         14 &        12 \\
GPXA-MW                           &          28 &         16 &        21 &       14 &         26 &        16 \\
GPXA-MT                           &          26 &         13 &        18 &       13 &         24 &        13 \\
GPXA-T                            &          30 &         12 &        17 &       10 &         27 &        12 \\
GPXC-W                            &          18 &         13 &        16 &       12 &         18 &        13 \\
GPXC-MW                           &          26 &         16 &        19 &       13 &         24 &        15 \\
GPXC-MT                           &          24 &         14 &        16 &       13 &         22 &        14 \\
GPXC-T                            &          47 &         10 &        20 &        9 &         41 &        10 \\
\hline
SPXA-W$^d$                        &          14 &         14 &        11 &       11 &         13 &        13 \\
SPXC-W                            &          19 &         14 &        13 &       11 &         18 &        13 \\
SPXC-MT                           &          30 &         15 &        17 &       13 &         27 &        14 \\
SPXA-W-H                          &          14 &         14 &        11 &       11 &         13 &        13 \\
SPXC-W-H                          &          19 &         14 &        13 &       11 &         18 &        13 \\
SPXC-MT-H                         &          30 &         15 &        17 &       13 &         27 &        14 \\
SPXA-W-HT                         &          14 &         14 &        11 &       11 &         13 &        13 \\
SPXC-W-HT                         &          19 &         14 &        13 &       11 &         18 &        13 \\
SPXC-MT-HT                        &          30 &         14 &        17 &       13 &         27 &        14 \\
\hline
CIDER-X-AHW$^e$                   &          14 &         14 &        11 &       10 &         13 &        13 \\
CIDER-X-CHW                       &          19 &         14 &        13 &       11 &         17 &        13 \\
CIDER-X-CHMT                      &          33 &         16 &        19 &       15 &         29 &        16 \\
\bottomrule
\end{tabular}\\
\end{center}
$^a$The first nine functionals are semi-local.
$^b$The next functional is a GP with semi-local descriptors ($x_1$ and $x_2$ from eq 19).
$^c$The next eight functionals (GPX...) are squared-exponential Gaussian processes, with the letter after X being the descriptor version and the code after the dash being the width (W=WIDE, MW=MEDIUM WIDE, MT=MEDIUM TIGHT, T=TIGHT).
$^d$The next nine functionals (SPX...) are spline-mapped additive GPs with different constraints (-H=homogeneous electron gas (HEG) limit, -HT=HEG and atomic tail limits).
$^e$The final three models are the CIDER functionals tested in the main text of the paper, which are constrained by the HEG but not the atomic tail limit. The difference between the SPX... and CIDER... functionals is that SPX functionals normalized $x_2$ using $\frac{2}{1+x_2}-1$, while the final CIDER functionals used $\frac{2}{1+{x_2^{}}^2}-1$.
\end{table}

The hyperparameters of the GP kernel were optimized using maximum-log-likelihood optimization with the L-BFGS algorithm in \textsc{scikit-learn}~\cite{Pedregosa2011}. The last remaining hyperparameters were the choice between Version A and C descriptors and the choice between WIDE, MEDIUM-WIDE, MEDIUM-TIGHT, and TIGHT length-scales for the Gaussian kernels. Combining these decisions gives eight possible models. Each descriptor set was used to train a GP with a squared-exponential kernel and then applied to predict the exchange energies on the Jensen validation set. As shown in Table \ref{tab:cider_val}, all of the models have good performance for the total exchange energy, but the models with shorter length-scales have worse atomization exchange energies, indicating that they memorize the isolated atoms poorly. However, using shorter length-scales is ideal because it will result in a linear-scaling implementation of CIDER being faster. Therefore, we selected the Version A WIDE descriptors and Version C WIDE and MEDIUM-TIGHT descriptors for further investigation.

It was then necessary to answer the question of whether the HEG limit affects the model performance at all. To do so, the three descriptor sets above were used to train additive GP models with and without the HEG limit. As shown in Table \ref{tab:cider_val}, the HEG limit does not affect model performance. Apparently, the HEG is too dissimilar from molecular charge densities to make a difference.

Lastly, we note that during these initial investigations, $x_2$ was normalized using $\frac{2}{1+x_2}-1$, rather than $\frac{2}{1+{x_2^{}}^2}-1$ as in eq 19. However, when we initially attempted SCF calculations with the former normalization approach, the exchange model was unstable; many calculations failed to converge. On the other hand, most DFT calculations performed with models that use the latter normalization are stable. This normalization was therefore used for models explored in the main text of the paper, but as shown in Table \ref{tab:cider_val}, the performance on static densities is not affected. We suspect that having $\pdv{F_x}{\alpha}\biggr\rvert_{\alpha=0}=0$ improves the stability of the model.

\subsection{Mapping Details}

The additive GPs were mapped to a cubic spline. For the MEDIUM-TIGHT descriptor model, the number of grid points in each direction was chosen by dividing 8 by the descriptor length-scale. For the two WIDE descriptor models, the same approach was done, except 8 was divided by the length-scale for $x_1$, and 6 was divided by the length-scale for all other descriptors. For each axis, a maximum number of grid points was set at 120. This approach ensured that the number of grid points along each axis was sufficient to describe the length-scale of the function while also preventing the grids from becoming too large.

\section{Additional Benchmarking of the Gaussian Process Models}

\subsection{Benchmarking Training Set Size}

\begin{figure}
    \centering
    \includegraphics[width=\textwidth]{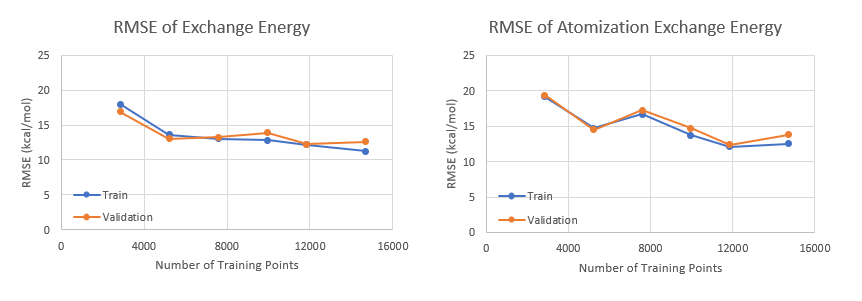}
    \caption{RMSE of Exchange Energy and Atomization Exchange Energy with respect to training set size on the Jensen validation set, for the feature set in eqs 17 and 19 and a squared-exponential (non-additive) kernel.}
    \label{fig:train_size}
\end{figure}

As mentioned in the main text, the training set size of 14,695 was chosen to maximize the training set size while maintaining tractable computational cost. In this section, we benchmarked models with different training set sizes on the feature vector A used in the main text (eqs 17 and 19). As shown in Figure \ref{fig:train_size}, the training set size has a weak effect on the validation set RMSE, with a slight decrease with increasing training set size. The RMSE for exchange energy is nearly identical for training set sizes of a third that used in the main text, so it might be possible to use a smaller training set in the future to benchmark models more quickly. However, the larger training set size used for the production models is clearly not detrimental to model accuracy.

\subsection{Benchmarking Choice of Feature Vector}

\begin{table}[]
\begin{center}
\caption{RMSE of atomization exchange energy (AEX) and total exchange energy (EX) for different model sizes in kcal/mol.}
\begin{tabular}{lrrrrrr}
\toprule
{} &  AEX Closed &  EX Closed &  AEX Open &  EX Open &  AEX Total &  EX Total \\
Functional$^a$ &             &            &           &          &            &           \\
\midrule
M0: 1-2   &          94 &        277 &        52 &      157 &         84 &       246 \\
M1: 1-3,9-10   &          49 &         30 &        33 &       23 &         45 &        28 \\
M2: 1-4,6,9-10 &          29 &         21 &        24 &       20 &         28 &        21 \\
M3: 1-8,11     &          45 &         23 &        35 &       20 &         42 &        22 \\
M4: 1-11       &          14 &         13 &        13 &       11 &         14 &        13 \\
M5: 1-10$^b$       &          14 &         13 &        13 &       11 &         14 &        13 \\
\bottomrule
\end{tabular}\label{tab:model_sizes}
\end{center}
$^a$The feature set indicates which features were used as input to the GP, with the features numbered as in eq 17 and eq \ref{eq:x11}. For example the model in the first row uses the feature vector $[x_1,x_2,x_3,x_9,x_{10}]$. All features use the WIDE parameters defined above.\\
$^b$This is the feature vector used in the models discussed in the main text.
\end{table}

Equation 17 of the main text lists the features used for the final models in this paper. As mentioned in the main text, the features contain a complete list of spherical harmonic contractions up to $l=2$, with the exception of
\begin{equation}
    x_{11}=C(\mathbf{G}_{01},\mathbf{G}_{02},\mathbf{G}_{01})\label{eq:x11}
\end{equation}
In addition to these contractions, two additional $l=0$ features are included, which have a longer ($x_9$) and shorter ($x_{10}$) length scale than the other descriptors, respectively. This choice of feature vector in eq 17 is not unique; one could raise or lower the maximum $l$ value of the functions being contracted, remove the additional $l=0$ features, append $x_{11}$ to the feature vector, etc. In this study, features with $l>2$ were excluded to keep computational cost reasonable.

To demonstrate that the choice of feature vector in eq 17 is reasonable, Table \ref{tab:model_sizes} shows the validation RMSE on the Jensen set for squared-exponential Gaussian Process models for different feature subsets. The models are specified as follows:
\begin{itemize}
    \item M0 contains only semi-local features
    \item M1 contains only $l=0$ features.
    \item M2 contains $l=0,1$ features.
    \item M3 contains all contractions of $l=0,1,2$ features but does not include the additional $l=0$ features $x_9$ and $x_{10}$.
    \item M4 contains all contractions of $l=0,1,2$ features along with $x_9$ and $x_{10}$.
    \item M5 is equivalent to eq 17.
\end{itemize}
M5 outperforms all the other models except M4, which has the same RMSE as M5 on the validation set. Therefore, $x_{11}$ does not improve performance (compare M4 and M5), while $x_9$ and $x_{10}$ are necessary to optimize the validation set accuracy (compare M3 and M5). Comparing M1 and M2 to M5 shows that only using $l=0$ and $l=1$ features is not sufficient to accurately describe the exchange energy.

\begin{figure}
    \centering
    \includegraphics[width=\textwidth]{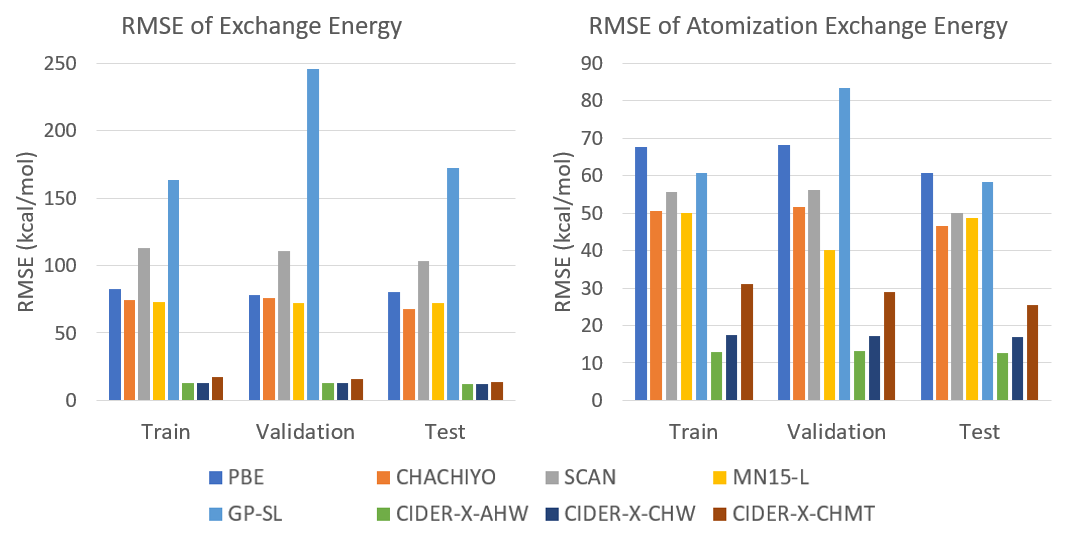}
    \caption{RMSE on Jensen dataset including the semi-local Gaussian Process model GP-SL. Same as Figure 1 in the main text with the addition of GP-SL.}
    \label{fig:slbench}
\end{figure}

Notably, the performance of the semi-local model, M0, is even worse than other typical semi-local models, as shown in Figure \ref{fig:slbench}, where the semi-local model is labeled GP-SL. We suspect this is because the model is trained to the non-unique exchange energy density (as opposed to total exchange energies), which can provide useful regularization for the more complex models but places burdensome constraints on the less expressive semi-local models. Therefore, any semi-local models or models with small feature sets trained using a CIDER-like approach might need to relax the objective of training to exchange energy densities in order to achieve reasonable accuracy.

\section{Linear Scaling Exchange Functional\label{si:linear_algo}}

The current implementation of the CIDER functional presented here scales as $\mathcal{O}(N^3)$ (due to the auxiliary basis transformation), but it also has the expensive rate-limiting step for projecting the descriptor basis onto the density auxiliary basis, which scales as $\mathcal{O}(N^2)$ and will dominate for small and medium molecules due to the large prefactor introduced by the real-space grid. This whole algorithm can actually be done in an asymptotically linear-scaling fashion per iteration with some integral pre-screening optimizations. There is a quadratic-scaling one-time setup step. Here is a sketch of said algorithm.

\begin{enumerate}
    \item Find the pairs of atomic orbitals $\ket{\mu\nu}$ with non-negligible density distribution ($\mathcal{O}(N^2)$). This results in $\mathcal{O}(N)$ pairs.
    \item Find which descriptor basis functions $\ket{g_i}$ have nonzero projection onto the auxiliary basis $\ket{\Theta_P}$ ($\mathcal{O}(N^2)$). This results in a linear number of projections $\braket{\Theta_P}{g_i}$. Note that because the descriptor basis length-scales change during the SCF loop as the density changes, doing this step once at the beginning of the SCF loop is inexact. However, if a little bit of extra buffer space is added to account for density fluctuations, this should be a good approximation. This step can also be performed once per iteration. This will make each iteration quadratic-scaling, but the prescreening step will probably not be the bottleneck for most systems.
    \item For each SCF iteration (each step can be made asymptotically linear-scaling with integral pre-screening):
    \begin{enumerate}
        \item Calculate the density in auxiliary space using $\braket{\Theta_P}{\mu\nu}$.
        \item Calculate the integrals $\int d^3\vec{r} n(\vec{r}) g_i(\vec{r})$ using $\braket{\Theta_P}{g_i}$.
        \item Calculate the XC energy from the above integrals.
        \item Calculate the energy derivatives with respect to $\ket{\Theta_P}$.
        \item Transform the energy derivatives into atomic orbital space using $\braket{\Theta_P}{\mu\nu}$ to get the XC potential.
    \end{enumerate}
\end{enumerate}

\section{Calculation of Exchange Energy Densities\label{si:xed}}

The exchange energy for spin channel $\sigma$ of a system is
\begin{equation}
    E_x^{\sigma} = -\frac{1}{2} \int \diff^3\mathbf{r}\,\diff^3\mathbf{r}'\, 
    \frac{|n_{1,\sigma}(\mathbf{r},\mathbf{r}')|^2}{|\mathbf{r}-\mathbf{r}\,'|}
    \label{eq:exchange_energy_dm}
\end{equation}
For brevity, the $\sigma$ index will be dropped for the remainder of this section. The above equation can also be expressed in terms of the occupied molecular orbitals (Kohn-Sham eigenstates) $\phi_i$ as
\begin{equation}
    E_x = -\frac{1}{2} \sum_{ij} \int \diff^3\mathbf{r}\,\diff^3\mathbf{r}'\, 
    \frac{|\phi_i(\mathbf{r})\phi_j(\mathbf{r}')|^2}{|\mathbf{r}-\mathbf{r}\,'|}\label{eq:ex_si}
\end{equation}

The products of orbitals can be expanded in a density fitting basis $\{\Theta_p(\mathbf{r})\}$~\cite{Pedersen2009}:
\begin{equation}
    \phi_i(\mathbf{r}) \phi_j(\mathbf{r}) = \sum_{p} C_{p}^{ij} \Theta_p(\mathbf{r})
\end{equation}
Using the above equation, eq \ref{eq:ex_si} can be transformed into the DF basis,
\begin{equation}
    E_x = -\frac{1}{2} \sum_{pq} D_{pq} \int \diff^3\mathbf{r}\,\diff^3\mathbf{r}'\,
    \frac{\Theta_p(\mathbf{r})\Theta_q(\mathbf{r}')}{|\mathbf{r}-\mathbf{r}\,'|}
\end{equation}
with $D_{pq}=\sum_{ij} C_{p}^{ij} C_{q}^{ij}$.

A valid definition of the exchange energy $e_x(\mathbf{r})$ is any definition that satisfies
\begin{equation}
    E_x=\int \diff^3\mathbf{r}\, e_x(\mathbf{r}).
\end{equation}
The most commonly discussed form of the exchange energy density is~\cite{Tao2003a}
\begin{equation}
    e_x^{\lambda}(\mathbf{r}) = -\frac{1}{2} \sum_{pq} D_{pq} \int \diff^3\mathbf{r}'\,
    \frac{\Theta_p(\mathbf{r}+\lambda\mathbf{r}')
    \Theta_q(\mathbf{r}-(1-\lambda)\mathbf{r}')}{|\mathbf{r}\,'|}
    \label{eq:lambda_exdens}
\end{equation}
with $\lambda=1$ defining the conventional exchange energy density, which is easily computed using standard analytical integration approaches because the $\Theta_q(\mathbf{r})$ term can be pulled out of the integral. Now suppose that the density basis takes the form of standard Gaussian-type orbital (GTO) primitives,
\begin{equation}
    \Theta_p=Y_{lm}(\mathbf{\hat{r}}) r^l e^{-\alpha(\mathbf{r}-\mathbf{A})^2}
\end{equation}
where $p=\{l_1,m_1,\alpha,\mathbf{A}\}$ and $q=\{l_2,m_2,\beta,\mathbf{B}\}$ index the GTOs. When the $\lambda$-dependent position coordinates are substituted into the above equation, the following relationships apply:
\begin{align}
    \Theta_{l_1,m_1,\alpha,\mathbf{A}}(\mathbf{r}+\lambda\mathbf{r}')
    &=\lambda^{l_1} \Theta_{l_1,m_1,\alpha',\mathbf{A}'}(\mathbf{r}')\\
    \mathbf{A}'&=\frac{\mathbf{A}-\mathbf{r}}{\lambda}\\
    \alpha'&=\alpha\lambda^2\\
    \Theta_{l_2,m_2,\beta,\mathbf{B}}(\mathbf{r}-(1-\lambda)\mathbf{r}')
    &=(\lambda-1)^{l_2} \Theta_{l_2,m_2,\beta',\mathbf{B}'}(\mathbf{r}')\\
    \mathbf{B}'&=\frac{\mathbf{r}-\mathbf{B}}{1-\lambda}\\
    \beta'&=\beta(1-\lambda)^2
\end{align}
With $p'=\{l_1,m_1,\alpha',\mathbf{A}'\}$ and $q'=\{l_2,m_2,\beta',\mathbf{B}'\}$, eq \ref{eq:lambda_exdens} can be rewritten as
\begin{equation}
    e_x^{\lambda}(\mathbf{r}) = -\frac{1}{2} \sum_{pq} D_{pq}
    \lambda^{l_1} (\lambda-1)^{l_2}
    \int d^3\mathbf{r}'\,
    \frac{\Theta_{p'}(\mathbf{r}')
    \Theta_{q'}(\mathbf{r}')}{|\mathbf{r}\,'|}\label{eq:lambda_exdens_gto}
\end{equation}
Note that $p'$ and $q'$ are functions of $\mathbf{r}$. The integrals in the above equation are standard 3-center 2-electron integrals with the third center a point charge distribution at $\mathbf{0}$. The molecular integral library \textsc{libcint}~\cite{Sun2015libcint} was edited to compute these integrals in a fork available at \url{https://github.com/kylebystrom/libcint}. A private development branch of \textsc{CiderPress} can be obtained via request to the corresponding author, which contains a module \verb+loc_analyzers+ that computes $e_x^{\lambda}$ for $\lambda\in[0.5,1.0)$. For $\lambda=1.0$, the exchange energy density should be computed using conventional approaches because eq \ref{eq:lambda_exdens_gto} has a singularity.

\begin{figure}[ht]
    \centering
    \includegraphics[width=\textwidth]{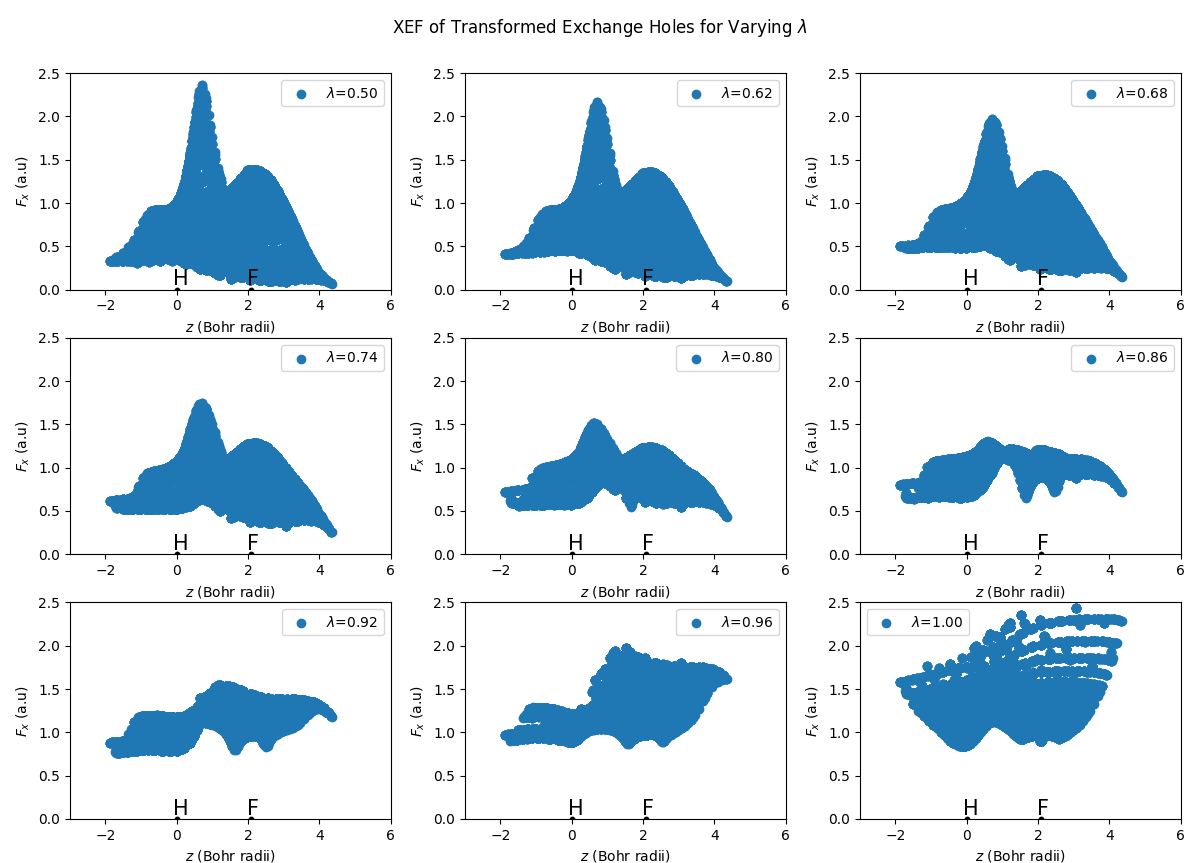}
    \caption{XEFs for varying values of $\lambda$ for the Hydrogen Fluoride molecule.}
    \label{fig:xed}
\end{figure}

Figure \ref{fig:xed} shows the Exchange enhancement factors (XEFs) for the Hydrogen Fluoride molecule for varying values of $\lambda$. The $\lambda=1$ (conventional) XEF is smooth but diverges at the atomic tails, while the highly localized $\lambda=0.5$ XEF has a large peak in the bonding region. The smallest range of the XEF is found for $\lambda=0.86$. We found that training to this XEF did not improve accuracy, but it might be a promising avenue for future research to further explore the use of different values of $\lambda$ for generating reference values for the XEF.

\section{Difference Between KS and HF Exchange Energies}

The following discussion explains the subtleties in the definition of the exact exchange energy for different mean-field electronic structure calculations, expanding on the note in the main text about the difference between the Kohn-Sham and Hartree-Fock exchange energies. It uses the formalism in the work by G\"orling and Ernzerhof~\cite{Gorling1995} on the difference between the HF and KS exchange energies.

In a self-consistent field (SCF) calculation with a single Slater determinant (Kohn-Sham DFT, Hartree-Fock, hybrid DFT, etc.), one self-consistently minimizes the energy functional $E[\Phi]$ of the system with respect to the set of occupied orbitals $\{\phi_i(\mathbf{r})\}$ of the single Slater determinant wavefunction $\Phi$. The two most simple examples of these energy functionals are the pure Kohn-Sham (KS) functional $E^{KS}[\Phi]$ and the Hartree-Fock (HF) functional $E^{HF}[\Phi]$:
\begin{align}
    E^{KS}[\Phi]&=\mel{\Phi}{\hat{T}}{\Phi}+V_{ext}[n] + U[n] + E_{x}[n]+E_{c}[n]\label{eq:supp_eks}\\
    E^{HF}[\Phi]&=\mel{\Phi}{\hat{T}+\hat{V}_{ee}}{\Phi}+V_{ext}[n]\label{eq:supp_ehf}
\end{align}
In these energy functionals, some terms, like the external potential energy $V_{ext}[n]$ and Hartree energy $U[n]$ are functionals only of the density $n(\mathbf{r})=\sum_i|\phi_i(\mathbf{r})|^2$, rather than the wavefunction $\Phi$. Others, like the kinetic energy $\mel{\Phi}{\hat{T}}{\Phi}$, are computed from the orbitals of the Slater determinant. For a given ground state density distribution $n(\mathbf{r})$, we need to minimize the energy to obtain the ground-state orbitals:
\begin{align}
    \Phi^{KS}[n]=\argmin_{\Phi\rightarrow n(\mathbf{r})} E^{KS}[\Phi]\label{eq:phiks}\\
    \Phi^{HF}[n]=\argmin_{\Phi\rightarrow n(\mathbf{r})} E^{HF}[\Phi]\label{eq:phihf}
\end{align}
In this case, we use the constrained search formalism of Levy and Lieb~\cite{Levy1979,Levy1982,Lieb1983}, meaning that the orbitals are constrained to yield a density distribution $n(\mathbf{r})$. While this is not how a practical SCF calculation is performed, it helps illustrate the relationships between the orbitals obtained in different methods by measuring them when the density is constrained to be the same for each method. It is also important for this paper because the CIDER functional is trained on densities which are self-consistently optimized using PBE, but not the CIDER functional itself. Therefore, the orbitals will be slightly different for GGAs, meta-GGAs, and hybrids, even for the same density, and this will affect the exchange energy.

Noting that terms which are explicit functionals of $n(\mathbf{r})$ only do not impact the above minimization, we can neglect them when we plug eqs \ref{eq:supp_eks} and \ref{eq:supp_ehf} into eqs \ref{eq:phiks} and \ref{eq:phihf}, respectively, yielding 
\begin{align}
    \Phi^{KS}[n]&=\argmin_{\Phi\rightarrow n(\mathbf{r})}\mel{\Phi}{\hat{T}}{\Phi}\\
    \Phi^{HF}[n]&=\argmin_{\Phi\rightarrow n(\mathbf{r})}\mel{\Phi}{\hat{T}+\hat{V}_{ee}}{\Phi}
\end{align}
As a result, the KS and HF orbitals are constructed by optimizing different objective functions. In practical SCF calculations, this means that the effective potential of the one-particle Hamiltonian is different for HF and KS approaches. In particular, the effective potential in the Kohn-Sham approach is of the local form $v_{eff}(\mathbf{r})$, while the Hartree-Fock effective potential is of the nonlocal form $v_{eff}(\mathbf{r},\mathbf{r}')$.

Following the above approach, one can establish more sophisticated energy functionals, such as the hybrid DFT energy functional $E^{hyb}[\Phi]$ and meta-GGA energy functional $E^{mGGA}[\Phi]$:
\begin{align}
    E^{hyb}[\Phi]&=\mel{\Phi}{\hat{T}+a\hat{V}_{ee}}{\Phi}+V_{ext}[n]+(1-a)(U[n]+E_x^{GGA}[n])+E_c^{GGA}[n]\\
    E^{mGGA}[\Phi]&=\mel{\Phi}{\hat{T}}{\Phi}+E_{xc}\left[n,\tau[\Phi]\right]+V_{ext}[n]+U[n]
\end{align}
The XC energy for the meta-GGA cannot be ignored during wavefunction optimization because the kinetic energy $\tau[\Phi](\mathbf{r})=\frac{1}{2}\sum_i|\nabla \phi_i(\mathbf{r})|^2$ is orbital-dependent. Optimizing the orbitals results in
\begin{align}
    \Phi^{hyb}[n]&=\argmin_{\Phi\rightarrow n(\mathbf{r})}\mel{\Phi}{\hat{T}+a\hat{V}_{ee}}{\Phi}\\
    \Phi^{mGGA}[n]&=\argmin_{\Phi\rightarrow n(\mathbf{r})}\left(\mel{\Phi}{\hat{T}}{\Phi}+E_{xc}\left[n,\tau[\Phi]\right]\right)
\end{align}
For any of these optimized Slater determinants $\Phi[n]$, one can compute the one-particle density matrix $n_1(\mathbf{r},\mathbf{r}')=\sum_i\phi_i^*(\mathbf{r})\phi_i(\mathbf{r}')$ and compute the exact exchange energy from eqs 27 and 29 in the main text. Because the density matrix depends on the orbitals, however, the exchange energy will be different for each method above, even for the same density distribution, because the orbitals for a given density distribution differ from method to method.

Because PBE, the functional used to generate the training densities and orbitals, is a semi-local, orbital-independent functional, its orbitals correspond to $\Phi^{KS}[n]$. B3LYP is a hybrid functional, corresponding to $\Phi^{hyb}[n]$ with $a=0.2$. Lastly, CIDER is a meta-GGA (including some non-local features), resulting in $\Phi^{mGGA}[n]$. As a result, when optimized self-consistently for a given density distribution $n(\mathbf{r})$, all three approaches yield different sets of orbitals for $\Phi$, resulting in a different exact exchange energy. For this reason, comparison of the exact exchange energies obtained from PBE orbitals, B3LYP orbitals, and B3LYP-CIDER orbitals is not exact. However, as mentioned in G\"orling and Erzerhof's work~\cite{Gorling1995} and in the main text, the energy differences for these sets of orbitals is expected to be quite small, so it is sensible to train CIDER on PBE densities/orbitals and then test by comparison to B3LYP calculations.

\printbibliography

%% file: body/introduction.tex
\section{Introduction}

Density functional theory (DFT)~\cite{Hohenberg1964,Kohn1965} is an indispensable tool in computational chemistry and materials science due to its combination of efficiency and accuracy. While an exact theory, the exchange-correlation (XC) functional describing quantum mechanical effects must be approximated in practice, and this approximation is the key limiting factor in the accuracy of DFT. Hundreds of approximate XC functionals have been developed in an attempt to calculate accurate chemical data~\cite{Mardirossian2017}, but significant progress remains to be made in the development of highly accurate XC functionals. Several reviews cover the shortcomings of current XC functionals~\cite{Verma2020,Su2017,Mardirossian2017,Yu2016Perspective,Cohen2012}, such as self-interaction error, over-delocalization, and inconsistent description of static correlation. 

The shortcomings of existing approximations have motivated the development of machine learning (ML) models for more accurate functionals~\cite{Kalita2021}. This concept was first introduced by Tozer et al.~\cite{Tozer1996} and more recently pioneered by Burke and coworkers~\cite{Snyder2012,Snyder2013,Li2016}. Some of these models use projections of the electron density or density matrix onto atom-centered basis sets as input to an ML model, resulting in an atomic decomposition of the the XC energy~\cite{Dick2020,doi:10.1021/acs.jctc.0c00872,Margraf2021}. These models are highly accurate across a small set of systems similar to those on which the model is trained, but they do not match the universality of most conventional XC functionals. For example, DeePKS~\cite{doi:10.1021/acs.jctc.0c00872} is only applicable to closed-shell organic molecules. Another recent approach uses a real-space grid and convolves the density to create features in real space~\cite{Lei2019descriptors,Nagai2020}. This approach has good transferability across chemical space and requires less training data than other ML functionals (as shown by Nagai et al.~\cite{Nagai2020}, who used a training set of only three molecules), but the accuracy achieved thus far is not consistently better than conventional semi-empirical XC functionals. For example, the NN-NRA by Nagai et al. is more accurate than M06-L~\cite{Zhao2006} and M06~\cite{Zhao2008} for main group ionization potentials, but less accurate for barrier heights~\cite{Nagai2020}.

The latter real-space approach is effectively an extension of the semi-empirical approach to functional design, in which a parametric functional is fit to experimental or theoretical training data. There is a trade-off between designing functionals that fit experimental data well and functionals that obey analytically known constraints on the exact functional~\cite{Perdew2005,Yu2016Perspective}. It has been shown that semi-empirical functionals have improved accuracy for energetic data, but breaking exact constraints can make the functional less universal and transferable, resulting in poor description of density distributions~\cite{Medvedev2017}. Exact constraints for ML functionals have not received much attention thus far. However, one study on the 1D kinetic energy functional showed that the uniform scaling constraint improved the learning curve of a kernel ridge regression-based functional~\cite{Hollingsworth2018}, and several constraints were enforced in a recent ML meta-GGA by Dick and Fern\'andez-Serra~\cite{Dick2021}. One can also explicitly train an ML model to reproduce the correct density; it has been shown that training on both energetic data and density distributions can yield ML functionals that accurately predict both properties~\cite{Nagai2020,doi:10.1021/acs.jctc.0c00872,Li2021,Kasim2021,Dick2021}.

Comparing the various approaches above, it seems that the atomic decomposition approach lacks the physical intuition and data efficiency to develop broadly transferable functionals for the entire periodic table, whereas existing real-space approaches lack the flexibility to describe systems to a high degree of accuracy while maintaining sensible physical behavior like accurate density distributions. Therefore, we seek to develop a framework for constructing descriptors of the density distribution that can be used to design accurate and transferable XC functionals with ML. Such features must describe variations in the density that correlate with the target XC energy, but they need not have an intuitive physical meaning because the ML algorithm can learn the relationship between the features and XC energy. Ideally, the features should enable an ML model trained on limited data to generalize across a broad range of chemistries.

For this work, we focus on the Kohn-Sham (KS) exchange functional~\cite{Gorling1995}
\begin{equation}
    E_x^{\text{ex}}[n]=\mel{\Phi^{\text{KS}}[n]}{\hat{V}_{ee}}{\Phi^{\text{KS}}[n]}-U[n]
\end{equation}
where $\Phi^{\text{KS}}[n]$ is the KS Slater determinant, $\hat{V}_{ee}=\sum_{i<j}r_{ij}^{-1}$ is the Coulomb repulsion operator, and $U[n]=\int \diff^3\mathbf{r}_1 \diff^3\mathbf{r}_2 \,n(\mathbf{r}_1) n(\mathbf{r}_2)\, r_{12}^{-1}$ is the classical Coulomb interaction.

There are several reasons for focusing on the exchange functional, rather than the full XC functional. First, compared to the full XC energy, it is much easier to collect training data for $E_x^{\text{ex}}[n]$ because it can be computed analytically from a KS density matrix. This makes it ideal for initial studies on real molecular systems.
Second, evaluating the exact exchange energy is usually the computational bottleneck of hybrid DFT calculations, in which a fraction of exact exchange is mixed into an otherwise semi-local XC functional. For periodic solids, hybrid DFT can be orders of magnitude more expensive than semi-local DFT and scales poorly with system size, which places limits on its applications~\cite{Lin2016}. There have been recent developments in the efficient computation of the exchange energy for periodic systems~\cite{Lin2016,Hu2017,Carnimeo_2019,Vinson2020}, but these approaches are still more costly than semi-local DFT. In addition, the linear-scaling approaches are only applicable to systems with band gaps, making them impractical for applications like catalysis on metal surfaces. An efficient ML exchange model could drastically reduce the computational cost of a calculation compared to hybrid DFT while preserving its accuracy. This is important because there are some technologically relevant systems, like semiconductor point defects~\cite{Lany2008,Freysoldt2014} and battery materials~\cite{He2019}, for which semi-local DFT is inaccurate but can be corrected by mixing an empirically tuned fraction of exact exchange to form a hybrid functional~\cite{Seo2015,Urban2016,Alkauskas2011}. Similar applications might be found in areas like heterogeneous catalysis \cite{Wellendorff2015}, where semi-local DFT is also frequently inadequate to describe the systems of interest. An ML exchange functional could make accurate studies of these systems faster and more practical.

Studying the exchange functional on its own is also important for developing a robust combined XC functional. The exchange and correlation energy errors tend to cancel each other because the exchange-correlation hole is more localized than the exchange or correlation holes on their own. However, this cancellation of error is neither controlled nor universal~\cite{Medvedev2017}. This means that in situations where such cancellation of error does not occur, such as self-interaction dominated systems, an otherwise accurate XC functional could fail unexpectedly. Having an exchange functional that is accurate on its own provides a baseline for ensuring physical behavior in these systems. Semi-local functionals do rely heavily on cancellation of error effects, so it is an open question what descriptors, if any, are capable of describing the shape of the exchange hole with sufficient accuracy and efficiency to replace the exact exchange operator. Knowing the answer to this question could assist future functional design.

To demonstrate that ML can be used to design transferable and accurate functionals, we introduce the Compressed scale-Invariant DEnsity Representation (CIDER)---a set of descriptors of the density distribution that are invariant under uniform scaling of the density (i.e., for $n_{\gamma}(\mathbf{r})=\gamma^3n(\gamma\mathbf{r})$, invariant with respect to $\gamma$)---and use it to train a Gaussian process for the exchange functional. This scale-invariance allows the exchange model to obey the uniform scaling rule $E_x[n_{\gamma}]=\gamma E_x[n]$. The resulting exchange functional requires only 119 training systems to be transferable across the first four rows of the periodic table. It outperforms all semi-local exchange functionals investigated here for predicting the total and atomization exchange energies, and it accurately reproduces main-group atomization energies when replacing the exact exchange contribution in B3LYP~\cite{Stephens1994}. It has good numerical stability, allowing it to be used in self-consistent field calculations with standard integration grids. The CIDER approach thus provides an initial step toward a robust and efficient alternative to approximating exact exchange.

The rest of the paper is structured as follows: Section \ref{sec:cider} introduces the CIDER formalism, and Section \ref{sec:gp} describes the Gaussian process models used with the CIDER descriptors to train exchange functionals. Section \ref{sec:methods} describes the methods for performing the DFT calculations and training the CIDER functional, and Section \ref{sec:results} contains the results and a discussion of the new functional's performance across a diverse set of chemistries. Finally, Section \ref{sec:conclusion} concludes with a summary of the findings.

%% file: body/cider.tex
\section{The CIDER Formalism\label{sec:cider}}

To develop an ML model for the exchange energy, an expressive set of nonlocal descriptors of the density must be used as input to the model. One could use a neural network (NN) to learn the features from the raw density distribution in real space, but training features this way is data intensive, with $10^5$-$10^6$ training points used in recent works~\cite{Ryabov2020,Zhou2019}. In addition, these NNs rely on a specific grid structure over which convolutions are performed, which could impede their use in realistic production calculations. Alternatively, one could project the density or density matrix onto atomic basis sets, as is done in NeuralXC and DeePKS \cite{Dick2020,doi:10.1021/acs.jctc.0c00872}, but these two models do not incorporate any physical constraints into the features, making it infeasible to incorporate exact constraints into the model itself. The difficulty of incorporating physical constraints and intuition into such models could limit their transferability and universality. A compromise between these two approaches is to design features based on nonlocal convolutions of the density, as done by Lei and Medford~\cite{Lei2019descriptors} and by Nagai et al.~\cite{Nagai2020}, and then use these features as input to an ML model. We seek to improve on this third approach by designing descriptors that are highly expressive and which also constrain the resulting ML model to known properties of the exact functional.

The most important constraint for the exchange energy is derived from the principle of uniform scaling. Consider a density distribution $n(\mathbf{r})$, and a scaled density
\begin{equation}
    n_{\gamma}(\mathbf{r})=\gamma^3n(\gamma\mathbf{r})
\end{equation}
Several important exact constraints can be written using this scaled density. They include the uniform scaling rules for the non-interacting kinetic energy $T_s[n]$ and exchange energy $E_x[n]$~\cite{Levy1985,Dreizler1990}:
\begin{align}
    T_s[n_{\gamma}]&=\gamma^2T_s[n]\\
    E_x[n_{\gamma}]&=\gamma E_x[n]\label{eq:exchange_uscale}
\end{align}
Equation \ref{eq:exchange_uscale} implies that the exchange energy can be written as
\begin{equation}
    E_x[n_{\gamma}] = -\gamma\left(\frac{3}{4}\right)\left(\frac{3}{\pi}\right)^{1/3} \int \diff^3\mathbf{r}\, F_x[n](\mathbf{r})\, n^{4/3}(\mathbf{r})\label{eq:n43relation}
\end{equation}
which is the form used by most semi-local exchange functionals. The constant in front of the integral is chosen so that $F_x=1$ for the homogeneous electron gas (HEG). The functional for a spin-unpolarized density $E_x[n]$ can be extended to the spin-polarized case using the spin-scaling rule~\cite{Oliver1979}:
\begin{equation}
    E_x[n_{\uparrow},n_{\downarrow}]=\frac{1}{2}(E_x[2n_\uparrow]+E_x[2n_\downarrow])\label{eq:spinscaling}
\end{equation}
Because eq \ref{eq:spinscaling} uniquely and simply defines the spin-polarized exchange energy from the spin-unpolarized exchange energy, the remainder of the discussion in this section refers to the spin-unpolarized case.

The exchange enhancement factor (XEF) $F_x[n](\mathbf{r})$ in eq \ref{eq:n43relation} is independent of $\gamma$, a property which will be referred to as \emph{scale-invariance}. Therefore, it is reasonable to predict that an ML model for $F_x[n]$ will learn more efficiently if the feature vector $\mathbf{x}$ used as input to the model is scale-invariant ($\mathbf{x}[n_{\gamma}](\mathbf{r})=\mathbf{x}[n](\gamma\mathbf{r})$). The conventional descriptors of the gradient $\nabla n$ and kinetic energy density $\tau=\frac{1}{2}\sum_i|\nabla\phi_i|^2$ satisfy these rules:
\begin{align}
    \mathbf{s} &= \frac{\nabla n}{2(3\pi^2)^{1/3}n^{4/3}}\\
    \alpha &= \frac{\tau-\tau_W}{\tau_0}
\end{align}
where $\tau_W=|\nabla n|^2/8n$ is the kinetic energy density of a single-orbital system and $\tau_0=(3/10)(3\pi^2)^{2/3}n^{5/3}$ is the kinetic energy density of the HEG. The descriptor $\alpha$ was first introduced in by Sun et al.~\cite{Sun2013}. While these descriptors are useful, they are semi-local, so they cannot fully encode the complex, nonlocal structure of the exchange functional. The challenge is therefore to construct a set of \emph{nonlocal} descriptors that are scale-invariant, which can be used for accurately training an ML model of the functional $E_x[n]$.

Nonlocality can be introduced to the features by performing convolutions on the density with a short-range kernel, as done in the Near-Region Approximation (NRA) by Nagai et al.~\cite{Nagai2020}. However, these features are not scale-invariant. To achieve scale-invariance, we use an approach similar to that developed by Janesko and co-workers for ``Rung 3.5'' semi-empirical functionals~\cite{Janesko2010,Janesko2013,Janesko2018}. Rung 3.5 functionals use the one-particle density matrix $n_1(\mathbf{r},\mathbf{r}')$:
\begin{equation}
    n_1(\mathbf{r},\mathbf{r}')=\sum_i f_i \phi_i(\mathbf{r})\phi_i(\mathbf{r}')\label{eq:rdm1},
\end{equation}
where $\phi_i(\mathbf{r})$ are the Kohn-Sham orbitals and $f_i$ are the occupation numbers (2 for the occupied orbitals in a spin-unpolarized system and 0 for unoccupied orbitals).The density matrix $n_1(\mathbf{r},\mathbf{r}')$ is projected onto a semi-local model for the density matrix at each point $\mathbf{r}$ to construct the Rung 3.5 energy density $e_{\text{Rung3.5}}[n](\mathbf{r})$:
\begin{equation}
    e_{\text{Rung3.5}}[n](\mathbf{r})=\int \diff^3\mathbf{r}'\, \frac{n_1(\mathbf{r},\mathbf{r}')n_1^{\text{model}}(\mathbf{r}'-\mathbf{r};\mathbf{r})}{|\mathbf{r}-\mathbf{r}'|}
\end{equation}
Importantly, the model density matrix $n_1^{\text{model}}(\mathbf{r}'-\mathbf{r};\mathbf{r})$ is \emph{position-dependent} in a way that gives $e_{\text{Rung3.5}}[n](\mathbf{r})$ convenient scaling properties. For example, if $n_1^{\text{model}}(\mathbf{r}'-\mathbf{r};\mathbf{r})$ is the exchange hole of the HEG, then $e_{\text{Rung3.5}}[n_{\gamma}](\mathbf{r})=\gamma^4 e_{\text{Rung3.5}}[n](\gamma\mathbf{r})$.

Following this approach, we introduce a scale-invariant set of integral descriptors to describe the density distribution around a point:
\begin{align}
    G_{nlm}(\mathbf{r}) &= \int \diff^3\mathbf{r}'\,\, g_{nlm}(\mathbf{r}';\mathbf{r})\, n(\mathbf{r}+\mathbf{r}\,')\label{eq:gnl}\\
    g_{nlm}(\mathbf{r}';\mathbf{r}) &= B_0^{3/2} \sqrt{4\pi^{l-1}} \left( \frac{8\pi}{3} \right)^{\frac{l}{3}} Y_{lm}(\mathbf{\hat{r}}') (|\mathbf{r}'|\sqrt{a})^{2n+l} \mathrm{e}^{-a|\mathbf{r}'|^2}
\end{align}
In the above equation, the functions $Y_{lm}(\mathbf{\hat{r}}')$ represent the real spherical harmonics, and the exponent $a$ is a function of $\mathbf{r}$ and a semi-local functional of the density:
\begin{equation}
    a[n](\mathbf{r}) = \pi\left(\frac{n}{2}\right)^{2/3} \left[B_0+C_0\left(\frac{\tau}{\tau_0}-1\right)\right]\label{eq:anl}
\end{equation}
where $B_0$ and $C_0$ are tunable constants satisfying $B_0\ge C_0>0$. Notable choices include $B_0=1$---in which case $G_{000}=2$ for the spin-unpolarized HEG---and $B_0=C_0=\frac{6}{5\pi}(6\pi^2)^{2/3}$---in which case $a$ is related to the exponent for a single Slater-type orbital density (see Appendix \ref{app:math_slater} for a proof):
\begin{equation}
    n(\mathbf{r})\propto \mathrm{e}^{-(\sqrt{2a})r}\label{eq:slater_orb}
\end{equation}
The $B_0=1$ case is important because the exchange hole integrates to -1 (-2 if summed over spin for a spin-unpolarized system), so for $B_0=1$, $g_{000}$ is approximately shaped like the HEG exchange hole and has the correct norm for the exchange hole for the HEG. This gives the feature a sensible shape and length-scale, and it also gives a known HEG reference value for that feature, allowing the HEG constraint to be enforced.

If $B_0=C_0$, $a$ is finite in atomic core regions but will vanish at the center of single bonds, where $\tau=0$. If $C_0=0$, $a$ will vanish at the atomic tails as the density goes to zero. In each case, the length scale of the Gaussian distribution becomes infinite, which is incompatible with the goal of a computationally efficient functional and is also poor inductive bias because the exchange hole is localized around the reference point. Therefore, $B_0$ must be greater than $C_0$, and $C_0$ must be positive.

In this work, we use the constant $A$ to define the length-scale, such that
\begin{align}
    B_0&=A\label{eq:b0a}\\
    C_0&=\frac{A}{32}\frac{6}{5\pi}(6\pi^2)^{2/3}\approx 0.18A
\end{align}
This choice was found to yield a smooth length-scale in real space. We also define several choices of $A$ for the model, resulting in different widths of the Gaussian kernels:
\begin{itemize}
    \item WIDE: $A=1$
    \item MEDIUM-WIDE: $A=2$
    \item MEDIUM-TIGHT: $A=4$
    \item TIGHT: $A=8$
\end{itemize}

These nonlocal features, along with the semi-local features $\mathbf{s}$ and $\alpha$, constitute the Compressed scale-Invariant DEnsity Representation (CIDER), so named because the features satisfy $G_{nlm}[n_{\gamma}](\mathbf{r})=G_{nlm}[n](\gamma\mathbf{r})$ and provide an efficient representation of the density distribution around a point $\mathbf{r}$ in real space.

The XC energy is invariant under rotation and translation of the system. The raw CIDER descriptors above are translationally invariant but not rotationally invariant. To create rotationally invariant descriptors, contractions using Clebsh-Gordon coefficients are performed. This is similar to the approaches used in the Tensor Field Network~\cite{Thomas2018}, the Moment Tensor Potential~\cite{Shapeev2016}, the RIDR functional~\cite{Margraf2021}, and Lei and Medford's nonlocal density features~\cite{Lei2019descriptors}, which are recently developed ML methodologies for chemistry and materials science.

In its current iteration, the CIDER descriptors are used to define the feature vector $\mathbf{x}$, where $\mathbf{G}_{nl}$ is a vector containing $G_{nlm}$ for $m\in\{-2l-1,-2l,...,2l+1\}$:
\begin{equation}
\begin{split}
    x_1 &= s^2\\
    x_2 &= \alpha\\
    x_3 &= G_{00}\\
    x_4 &= \norm{\mathbf{G}_{01}}^2\\
    x_5 &= \norm{\mathbf{G}_{02}}^2/\sqrt{5}\\
    x_6 &= \mathbf{s} \cdot \mathbf{G}_{01}\\
    x_7 &= C(\mathbf{s}, \mathbf{G}_{02}, \mathbf{s})\\
    x_8 &= C(\mathbf{s}, \mathbf{G}_{02}, \mathbf{G}_{01})\\
    x_9 &= G_{00}(2^{-4/3}A)\\
    x_{10} &= G_{00}(2^{4/3}A)
\end{split}
\label{eq:feature_vector}
\end{equation}
$G_{00}(\lambda A)$ indicates that the length-scale parameter $A$ was changed to $\lambda A$ for this integral. $C(\mathbf{a},\mathbf{b},\mathbf{c})$ contracts the two $l=1$ terms $\mathbf{a,c}$ and the $l=2$ term $\mathbf{b}$ to an $l=0$ term using Clebsh-Gordon coefficients (see Appendix \ref{app:math_cg} for details). For length-scale parameter A, eq \ref{eq:feature_vector} contains all possible rotationally invariant contractions of features with $l\le 2$, with the exception of $x_{11}=C(\mathbf{G}_{01},\mathbf{G}_{02},\mathbf{G}_{01})$. We found that this set of features provides a good balance between computational efficiency and model expressiveness. In principle, one could expand the feature set by including contractions of features with $l>2$ and by using multiple different length-scale parameters, but to keep the computational cost of feature evaluation tractable, this is not done here. One could also attempt to learn on smaller feature sets, including a semi-local model containing only $x_1$ and $x_2$. However, as discussed in the Supporting Information (Section S2.2), this results in insufficiently accurate models. In Section S2.2, we also show that adding the feature $x_{11}$ does not improve the accuracy, and excluding the additional $l=0$ features $x_9$ and $x_{10}$ is detrimental to accuracy. We therefore conclude that eq \ref{eq:feature_vector} is a reasonable choice of feature vector.

The above descriptors will be referred to as Version A. We also tried introducing $G_{nlm}$ descriptors for $n\ne 0$, yielding Version C (Compact) descriptors, which are the same as Version A except for the last two:
\begin{equation}
\begin{split}
    x_9 &= G_{10}\\
    x_{10} &= G_{00}(2A)
\end{split}
\label{eq:feature_vector_c}
\end{equation}
While $x_{10}$ is a ``tight'' descriptor like in Version A, $x_9$ has the same exponent as the other nonlocal features but is multiplied by $ar^2$. This version allows us to examine the effects of eliminating the widest feature (thus ``Compact''), which is useful because shorter-range features might be more amenable to computationally efficient evaluation in optimized implementations of the model.

Normalizing features is generally helpful in machine learning applications. Using the descriptors developed by Becke \cite{Becke1997}, Becke and Edgecombe~\cite{Becke1990}, and Mardirossian and Head-Gordon \cite{Mardirossian2015} as guidance, we apply the following transformations to the above descriptors:
\begin{equation}
\begin{split}
    x_1 &\rightarrow \frac{\gamma_x x_1}{1 + \gamma_x x_1}\\
    x_2 &\rightarrow \frac{2}{1+{x_2^{}}^2}-1\\
    x_3 &\rightarrow \frac{\gamma_{0a} x_3}{1 + \gamma_{0a} x_3} - c_3\\
    x_4 &\rightarrow \frac{\gamma_1 x_4}{1 + \gamma_{1} x_4}\\
    x_5 &\rightarrow \frac{\gamma_2 x_5}{1 + \gamma_{2} x_5}\\
    x_6 &\rightarrow x_6\sqrt{\frac{\gamma_x}{1 + \gamma_x x_1}} \sqrt{\frac{\gamma_1}{1 + \gamma_{1} x_4}}\\
    x_7 &\rightarrow x_7\frac{\gamma_x}{1 + \gamma_x x_1} \sqrt{\frac{\gamma_2}{1 + \gamma_{2} x_4}}\\
    x_8 &\rightarrow x_8 \sqrt{\frac{\gamma_x}{1+\gamma_x x_1}} \sqrt{\frac{\gamma_1}{1+\gamma_1 x_4}} \sqrt{\frac{\gamma_2}{1+\gamma_2 x_5}}\\
    x_9 &\rightarrow \frac{\gamma_{0b} x_9}{1 + \gamma_{0b} x_9} - c_9\\
    x_{10} &\rightarrow \frac{\gamma_{0c} x_{10}}{1 + \gamma_{0c} x_{10}} - c_{10}
\end{split}
\label{eq:feature_vector_norm}
\end{equation}
The resulting transformed descriptors all fall in a finite range, making them more convenient for ML models. The hyperparameters in the equations above were selected heuristically as described in the Supporting Information (Section S1), but they could also be optimized, if desired, by treating them as hyperparameters of Gaussian process regression models, as explained in Section \ref{sec:gp}. The constants $c_3,c_9,c_{10}$ guarantee that the zero feature vector $\mathbf{x}=\mathbf{0}$ corresponds to the HEG. The HEG limit can therefore be enforced by setting $F_x=1$ for $\mathbf{x}=0$ in eq \ref{eq:n43relation}.

%% file: body/gps.tex
\section{Gaussian Process Exchange Models\label{sec:gp}}

To train ML models of the functionals, we employ Gaussian processes (GPs), which are commonly used for non-parametric regression in Bayesian statistical learning models~\cite{Rasmussen2005}. For a training set size $N$, matrix of inputs $\mathbf{X}$ (the set of feature vectors $\mathbf{x}^{(i)}$ for the training points $i=\{1,...,N\}$), vector of outputs $\mathbf{y}$, and kernel function $k(\mathbf{x},\mathbf{x}')$, the standard GP relations for the predictive mean $f(\mathbf{x})$ and variance $\sigma^2$ are
\begin{align}
    f(\mathbf{x}_*)&=\mathbf{k}_*^\top(\mathbf{K}+\sigma_{\text{noise}}^2\mathbf{I})^{-1}\mathbf{y}\label{eq:gp}\\
    \sigma^2(\mathbf{x}_*)&=k(\mathbf{x}_*,\mathbf{x}_*)-\mathbf{k}_*^\top(\mathbf{K}+\sigma_{\text{noise}}^2\mathbf{I})^{-1}\mathbf{k}_*
\end{align}
In the above equation, $\mathbf{x}_*$ is the test point; $\mathbf{k}_*$ is a vector containing $k(\mathbf{x}^{(i)},\mathbf{x}_*)$ for each $\mathbf{x}^{(i)}$ in $\mathbf{X}$; $\mathbf{I}$ is the identity matrix; and $K_{ij}=k(\mathbf{x}^{(i)},\mathbf{x}^{(j)})$ for each $\mathbf{x}^{(i)},\mathbf{x}^{(j)}$ in $\mathbf{X}$. $k(\mathbf{x},\mathbf{x}')$ can be any function satisfying the rules for an inner product, and it may contain a set of hyperparameters $\theta$ that may need to be optimized. These hyperparameters can be chosen by maximizing the marginal likelihood
\begin{equation}
    \ln p(\mathbf{y}|\mathbf{X},\theta)=-\frac{1}{2}\mathbf{y}^\top(\mathbf{K}+\sigma_{\text{noise}}^2\mathbf{I})^{-1}\mathbf{y}-\frac{1}{2}\ln|\mathbf{K}+\sigma_{\text{noise}}^2\mathbf{I}|-\frac{N}{2}\ln 2\pi
\end{equation}
The noise parameter $\sigma_{\text{noise}}$ can also be optimized in this manner.
One popular covariance kernel is the squared-exponential kernel
\begin{equation}
    k(\mathbf{x},\mathbf{x}')=\exp\left(-\frac{1}{2}|(\mathbf{x}-\mathbf{x}')\odot \mathbf{a}|^2\right)\label{eq:rbf}
\end{equation}
where $\odot$ represents the element-wise product, and $\mathbf{a}$ is a hyperparameter vector containing the inverse standard deviation in each direction in feature space. This kernel is used for some of the preliminary models discussed in the Supporting Information (Section S1) and as the base kernel for the additive kernel discussed below. The squared-exponential kernel yields highly expressive, nonlinear models because it is equivalent to linear regression on an infinite set of nonlinear basis functions~\cite{Rasmussen2005}. It is also is smooth and infinitely differentiable, which is important for ML functional models because they must be differentiated to obtain the exchange potential.

One drawback of the GP is that evaluating eq \ref{eq:gp} for a single test point scales linearly with the training set size $N$, so evaluating GPs with large training sets is computationally expensive. To enable efficient evaluation of the model, some developers of GP-based molecular dynamics force fields map the GP to a cubic spline of the features, so evaluation of the model has a low cost that is  independent of training set size~\cite{Glielmo2018,Xie2021BayesianStanene,Vandermause2020,Vandermause2021}. However, this approach is only feasible for $n \le 4$ features because the amount of memory required for the spline coefficients scales as $G^n$, with $G$ the grid size in each dimension. With the 10 features in the CIDER model, eq \ref{eq:rbf} cannot be mapped to a cubic spline.

This problem can be solved by the additive kernel developed by Duvenaud et al.~\cite{Duvenaud2011}:
\begin{equation}
    k_n^{\text{add}}(\mathbf{x},\mathbf{x}')=\sigma_n^2\sum_{1\le i_1<i_2<\cdots <i_n\le D} \prod_{d=1}^n k_{i_d}(x_{i_d},x_{i_d}')\label{eq:addrbf}
\end{equation}
where $D$ is the dimensionality of the feature vector, $n$ is the order of the kernel, and $k_i(x_i,x_i')$ is the kernel for feature $i$. For this kernel, the predictive mean can be linearly decomposed into terms with a maximum of $n$ features per term. Each of these terms can be mapped to a cubic spline in a memory-efficient manner for $n\le 4$. See Appendix \ref{app:spline_map} for a more detailed explanation.

\subsection{Model Details}

The additive Gaussian process approach allows interactions of any order in the descriptors to be accounted for by changing $n$ in eq \ref{eq:addrbf}. Using the feature vector $\mathbf{x}$ of eq \ref{eq:feature_vector}, the kernel for the exchange matrix is constructed by multiplying the $x_1$ kernel by an $n=2$ additive kernel for the other 9 descriptors:
\begin{equation}
\begin{split}
    k_{\text{CIDER}}(\mathbf{x},\mathbf{x}') =& \sigma_1^2 k_1(x_1,x_1')\\
    +& \sigma_2^2 k_1(x_1,x_1') \sum_{i=2}^{10} k_i(x_i,x_i')\\
    +& \sigma_3^2 k_1(x_1,x_1') \sum_{i=2}^{9} \sum_{j=i+1}^{10} k_i(x_i,x_i') k_j(x_j,x_j')
\end{split}
\label{eq:arbf_exchange}
\end{equation}
where $k_i(x_i,x_i')=\exp\left(-\frac{1}{2}a_i^2(x_i-x_i')^2\right)$ is a squared-exponential kernel. The exponent hyperparameters $a_i$ and weights $\sigma_n$ are fit using iterative maximum likelihood optimization. In practice, $\sigma_1$ and $\sigma_2$ vanish during hyperparameter optimization. The resulting predictive mean is mapped onto a sum of cubic splines, as described in previous work on molecular dynamics potentials~\cite{Glielmo2018,Xie2021BayesianStanene}. Because the maximum number of features per term in eq \ref{eq:arbf_exchange} is 3, the cubic splines are at most three-dimensional.

We train our predictive function $F_{x}^{\text{pred}}(\mathbf{x}[n](\mathbf{r}))$ to the XEF, $F_{x}[n](\mathbf{r})$, and evaluate $E_x[n]$ via eq \ref{eq:n43relation}:
\begin{align}
    F_{x}[n](\mathbf{r})&=\frac{e_{x}^{\text{ex}}[n](\mathbf{r})}{e_{x}^{\text{LDA}}(n(\mathbf{r}))}\\
    e_{x}^{\text{ex}}[n](\mathbf{r})&=-\frac{1}{4} \int \diff^3\mathbf{r}'\,
    \frac{|n_1(\mathbf{r},\mathbf{r}')|^2}{|\mathbf{r}-\mathbf{r}'|}\label{eq:exdens}\\
    e_{x}^{\text{LDA}}(n)&=-\frac{3}{4}\left(\frac{3}{\pi}\right)^{1/3} n^{4/3}
\end{align}
In eq \ref{eq:exdens}, $e_{x}^{\text{ex}}[n](\mathbf{r})$ is the exact exchange energy density, and $n_1(\mathbf{r},\mathbf{r}')$ is the density matrix (eq \ref{eq:rdm1}). Equation \ref{eq:exdens} is not a unique definition, as any $e_{x}^{\text{ex}}[n](\mathbf{r})$ for which
\begin{equation}
    E_{x}[n]=\int \diff^3\mathbf{r}\, e_{x}^{\text{ex}}[n](\mathbf{r})
\end{equation}
is equally valid, but eq \ref{eq:exdens} is the easiest to compute and was found to be the easiest to which to fit the model as well. The Supporting Information (Section S4) includes a discussion of alternative exchange energy densities. One of the challenges with the definition in eq \ref{eq:exdens} is that the XEF increases exponentially at the atomic tails. To make the model easier to train, the Chachiyo GGA exchange functional~\cite{Chachiyo2020}, which reproduces $F_x$ at atomic tails, was used as a baseline functional, and the difference $F_{x}[n](\mathbf{r})-F_{x}^{\text{Cachiyo}}[n](\mathbf{r})$ was learned.

The standard GP model uses a single noise parameter $\sigma_{\text{noise}}$, which is constant for all observations. However, we expect observations of $F_x$ to have a larger uncertainty for smaller densities because the exchange energy density is divided by $(n(\mathbf{r}))^{4/3}$. Therefore, we use a heteroskedastic noise model, in which each training point has a different noise parameter. This noise parameter is a function of the density $n(\mathbf{r})$ and takes the form
\begin{equation}
    \sigma_{\text{noise}}^2(n)=v_1+\frac{v_2}{1+t_2n}+\frac{v_3}{1+t_3n}\label{eq:dnoise}
\end{equation}
where $t_2=50$, $t_3=10^6$, and $v_1$, $v_2$, and $v_3$ are treated as hyperparameters and optimized by marginal likelihood maximization.

For numerical stability, the ML part of the functional is cut off at low densities. This is done using the function
\begin{equation}
    F_x=F_x^{\text{Chachiyo}}+f_{cut}F_x^{ML},
\end{equation}
where
\begin{align}
    f_{cut}&=\begin{cases}
    0 & n < n_{c,min} \\
    \frac{1}{2}(1-\cos(\pi x_{cut})) & n_{c,min} \le n \le n_{c,max} \\
    1 & n > n_{c,max}
    \end{cases} \label{eq:fcut} \\
    x_{cut}&=\frac{\ln(n_{c}/n_{c,min})}{\ln(n_{c,max}/n_{c,min})},\\
    n_{c}&=\max(n,n_{c,max}),
\end{align}
where $n_{c,max}=10^{-3}$ and $n_{c,min}=10^{-6}$ Bohr$^{-3}$.

Also for numerical stability, we define a value $a_{\text{cut}}$ below which the exponent $a(\mathbf{r})$ in eq \ref{eq:anl} is exponentially damped so that it cannot go to zero:
\begin{equation}
    a\rightarrow\begin{cases}
    a & a \ge a_{\text{cut}} \\
    a_{\text{cut}}\exp(a/a_{\text{cut}}-1) & a < a_{\text{cut}}
    \end{cases}
\end{equation}
$a_{\text{cut}}$ is set to $A/16$ for $A$ in eq \ref{eq:b0a}. This damping of the exponent violates the uniform scaling rule, but only at low densities for very diffuse orbitals.

%% file: body/methods.tex
\section{Methods\label{sec:methods}}

\subsection{Computational Details}

The \textsc{PySCF}~\cite{Sun2018} code was used for all calculations. The CIDER implementation is available in the \textsc{CiderPress} repository~\cite{ciderpress}. To compute the CIDER descriptors, we first projected the density matrix onto the def2 Coulomb fitting auxiliary basis~\cite{Weigend2006}. Then, the overlaps of the descriptor functions $g_{nlm}(\mathbf{r}';\mathbf{r})$ with the auxiliary basis were computed using the \textsc{libcint}~\cite{Sun2015libcint} library as interfaced through \textsc{PySCF}. These overlaps were contracted with the density to give the descriptors $G_{nlm}(\mathbf{r})$, which were then used to compute the XEF. Then, the functional derivatives were computed and used to evaluate the exchange matrix elements. The functional was treated self-consistently except for the $f_{cut}$ function in eq \ref{eq:fcut}, which was not differentiated with respect to the density. The formulas for the functional derivatives and matrix elements are written in Appendix \ref{app:derivs}. The \textsc{fireworks}~\cite{Jain2015} package was used to automate calculation workflows.

\subsection{Training and Testing the Gaussian Process}

The dataset developed by Jensen et al.~\cite{Jensen2017Data} was used to train, validate, and test the exchange model. This dataset is an expansion of the G2/97 test set~\cite{Schmider1998} that includes elements that are under-represented in the G2/97 set as well as six non-bonded systems~\cite{Jensen2017Paper}. The molecules were categorized by the elements they contained (Al, Be, B, Li, Mg, Na, P, Si, S, F, Cl), with the earlier elements in the list taking precedence (e.g., \ce{SF6} was categorized under S). The last set contained mostly organic molecules with none of these elements. Each sub-group was shuffled, and then for each dataset of size $N$, $N_{\text{train}}=\text{floor}(0.4N)$, $N_{\text{val}}=\text{round}(0.2N)$, and $N_{\text{test}}=N-N_{\text{train}}-N_{\text{val}}$ molecules were placed in the training, validation, and test sets, respectively. \ce{He2}, \ce{Be2}, and \ce{Ar2} were placed in the training, validation, and test sets, respectively. Later it was realized that the dataset contained a duplicated \ce{P2} molecule. Both duplicates were placed in the test set, so one was simply removed. The final partitions contained 79 training set molecules (55 closed-shell, 24 open-shell), 42 validation set molecules (29 closed-shell, 13 open-shell), and 90 test set molecules (64 closed-shell, 26 open-shell). In addition to these molecules, the training set was augmented with isolated atoms H-Kr, as well as the excited spin states of Sc, Ti, V, and Cr with $2S_z=3,4,5,4,$ respectively. The inclusion of isolated atoms helped better describe the shapes of the tails of atoms, and it also introduced some transition metal atoms to the training set. There were no transition metal-containing systems in the Jensen dataset.

The density matrices and energies for each system were evaluated self-consistently using PBE~\cite{Perdew1996} in the def2-QZVPPD basis set. From the PBE density matrix, the conventional exchange energy density (eq \ref{eq:exdens}) was calculated in real-space for each molecule on the level-3 grids implemented in \textsc{PySCF}. The same level-3 grids were used for all SCF calculations. The $S_z=0$ systems were performed in the Restricted Kohn-Sham (RKS) formalism, and the rest were performed in the unrestricted Kohn-Sham (UKS) formalism. From the real-space grids, 14,695 points were sampled from which to train the exchange enhancement factor model:
\begin{itemize}
    \item 481 points from closed-shell atoms,
    \item 4471 points from open-shell atoms,
    \item 6903 points from closed-shell molecules in the Jensen train partition,
    \item 2840 points from open-shell molecules in the Jensen train partition.
\end{itemize}
The training set size of 14,695 was chosen to maximize the training set size while keeping the computational cost tractable. For comparison, we benchmarked the train and validation performance of the CIDER functional for different training set sizes in Supporting Information Section S2.1 and found only a weak dependence of the accuracy on training set size. The points were selected randomly from the set of points for which the electron density was greater than $10^{-6}$ Bohr$^{-3}$. Finally, the HEG limit was appended as an additional training point, with the density set to $10^8$ to minimize the uncertainty in Equation \ref{eq:dnoise}. In summary, the CIDER model was trained to the exact exchange energy densities of density matrices obtained from self-consistent PBE calculations.

Gaussian process models with the kernel specified in eq \ref{eq:arbf_exchange} were trained to the training set described above. The parameters for the kernels were selected as described in detail in the Supporting Information (Section S1). Several combinations of exact constraints, descriptor types, and descriptor length-scales were tested, but we focused on three for the main body of this work:
\begin{itemize}
    \item CIDER-X-AHW: Version A descriptors, Homogeneous Electron Gas (HEG) constraint, WIDE descriptors ($A=1$ in eq \ref{eq:b0a}),
    \item CIDER-X-CHW: Same as CIDER-X-AHW but with Version C descriptors,
    \item CIDER-X-CHMT: Same as CIDER-X-CHW but with MEDIUM-TIGHT descriptors ($A=4$ in eq \ref{eq:b0a}).
\end{itemize}
All models were implemented in the \textsc{scikit-learn} package~\cite{Pedregosa2011}; the additive RBF was implemented as a custom \verb+Kernel+ object. The additive kernel models were mapped to cubic splines using the \textsc{interpolation.py} package~\cite{interpolation}. The best-performing model on the validation set (CIDER-X-AHW) was selected as the final model.

To evaluate the accuracy and transferability of CIDER-X-AHW on static densities, the exchange energies and atomization energies were evaluated on the test set PBE density matrices using CIDER-X-AHW. To evaluate the self-consistent performance of the functional, the atomization energies of the Jensen set were computed self-consistently using two functionals: CIDER-X-AHW alone (i.e. replacing exact exchange with CIDER in an HF calculation) and B3LYP-CIDER, in which the 20\% HF exchange contribution of the B3LYP hybrid functional~\cite{Stephens1994} was replaced with 20\% CIDER-X-AHW.

\subsection{Benchmarking CIDER on the Minnesota Database}

To evaluate the applicability of the CIDER functional to different properties, the B3LYP-CIDER functional was applied to the BH76 (barrier heights), IP23 (ionization potentials), and TMBE33 (transition metal bond energies) datasets from the Minnesota 2015B Database~\cite{Yu2016MN15} as distributed in the ACCDB collection of databases~\cite{Morgante2019}.

The Minnesota Database contains some systems, especially metal-containing systems, which can settle into metastable densities due to the presence of symmetry-breaking ground states. To avoid these stable minima, all Minnesota Database calculations were performed in the UKS formalism. For these calculations, each system was initially converged with a PBE calculation, and its internal stability was then tested using the \verb+uhf_internal+ tool in \textsc{PySCF}~\cite{Sun2018}. If the test found an instability and returned new orbitals, these orbitals were used to initialize another UKS calculation. This was repeated until a stable ground state was found. Finally, an SCF calculation was performed with the functional of interest starting from the stable ground state of the PBE calculation. All calculations were performed in the def2-QZVPPD basis~\cite{Weigend2003,Weigend2005,Rappoport2010}, with the def2 effective core potential (ECP)~\cite{Andrae1990} used for elements of Period 5 and above. Convergence of calculations was attempted to as low a threshold as $10^{-9}$ hartree (Ha), but calculations were accepted if they converged to $10^{-6}$ Ha.

To compare the non-relativistic DFT calculations performed here with the relativistic reference values in the Minnesota Database, the spin-orbit corrections (SOC) from Supplementary Information Table S19 of the revM06 paper~\cite{Wang2018} were applied.

%% file: body/results.tex
\section{Results and Discussion\label{sec:results}}

\begin{figure}[ht]
    \centering
    \includegraphics[width=\textwidth]{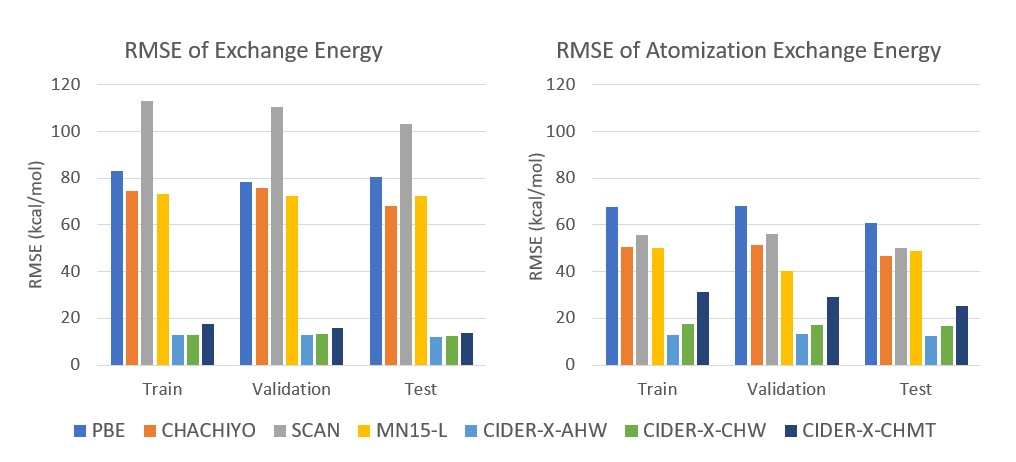}
    \caption{RMSE of the Kohn-Sham exchange energy and atomization exchange energy for semi-local functionals and CIDER functionals on the Train, Validation, and Test partitions of the Jensen dataset, in kcal/mol.}
    \label{fig:cider_jensen_static}
\end{figure}

\subsection{Static density benchmark}

As shown in Figure \ref{fig:cider_jensen_static}, three different versions of the CIDER functional all outperform existing semi-local functionals for predicting the exact exchange energy of PBE density matrices. We chose to compare with PBE~\cite{Perdew1996} and SCAN~\cite{Sun2015SCAN} because of their popularity, with the Chachiyo GGA~\cite{Chachiyo2020} because it is the baseline for our ML model, and with MN15-L~\cite{Yu2016MN15L} because it had the lowest root mean square error (RMSE) of the semi-local functionals we investigated for atomization exchange energy on the validation set. At first, this is surprising because MN15-L is a non-separable meta-generalized gradient approximation (mNGA), meaning that its exchange part does not obey the uniform scaling rule and therefore contains some correlation-like character~\cite{Yu2016MN15L}. Due to the cancellation of error between the exchange and correlation parts, however, it makes sense that the nonseparable XC part of MN15-L might achieve a lower error for exact exchange than other semi-local functionals. The Supporting Information (Table S1) contains the validation set errors for several other semi-local functionals~\cite{Becke1988Functional,Becke1989,Proynov2008,Tao2016,Wellendorff2014}, which perform similarly to the functionals discussed here.

The best-performing CIDER functional in Figure \ref{fig:cider_jensen_static} is CIDER-X-AHW, with an RMSE on the test set of 12.0 kcal/mol for exchange energy (EX) and 12.5 kcal/mol for atomization exchange energy (AEX), which is defined as the difference between EX and the value of EX for the molecule's constituent atoms. These errors are only 18\% (EX) and 27\% (AEX) of the RMSE of the most accurate semi-local functional, the Chachiyo GGA~\cite{Chachiyo2020}. Notably, the performance of all CIDER functionals is nearly identical on the train, validation, and test sets, indicating good transferability.

The other notable finding in Table \ref{fig:cider_jensen_static} is that longer length-scale descriptors (CIDER-X-AHW) do not improve the description of EX over shorter length-scale descriptors (CIDER-X-CHMT), but they do give an improvement for AEX. In particular, the accuracy of AEX degrades as the length-scale gets shorter, while the accuracy of EX stays about the same. This suggests that a descriptor with a short length-scale cannot describe the localized atomic density and the more delocalized molecular density simultaneously. On the other hand, descriptors with longer length-scales can accurately describe both single-center and multi-center exchange holes. Because of its accuracy on the validation set for both EX and AEX, CIDER-X-AHW was used for the self-consistent field calculations below. The adjustable parameters for eqs \ref{eq:feature_vector_norm} and \ref{eq:dnoise} for CIDER-X-AHW are $\gamma_{0a}=\gamma_{0b}=\gamma_{0c}=1/2$, $\gamma_1=0.025$, $\gamma_2=0.015$, $v_1=10^{-6}$, $v_2=0.000503$, and $v_3=0.391$.

\subsection{Self-consistent field calculations with B3LYP-CIDER}

\begin{table}[ht]
    \centering
    \begin{tabular}{|l|rr|rr|}
        \hline
        Partition & \multicolumn{2}{c|}{CIDER-X-AHW} & \multicolumn{2}{c|}{B3LYP-CIDER} \\
         & MAE & RMSE & MAE & RMSE \\
        \hline
        Train & 8.7 & 12.5 & 1.7 & 2.5 \\
        Validation & 9.0 & 12.9 & 1.8 & 2.5 \\
        Test & 7.9 & 11.6 & 1.6 & 2.3 \\
        \hline
        Combined & 8.4 & 12.2 & 1.7 & 2.4 \\
        \hline
    \end{tabular}
    \caption{MAE and RMSE of the CIDER-X-AHW and B3LYP functionals on the Jensen dataset in kcal/mol, with the reference values being HF for CIDER-X-AHW and B3LYP for B3LYP-CIDER.}
    \label{tab:cider_jensen_scf}
\end{table}

To be useful, the CIDER exchange functional must not only outperform semi-local exchange functionals, but also accurately match the results of calculations performed using HF or hybrid functionals. To test this, SCF calculations were performed using CIDER-X-AHW and B3LYP-CIDER. Table \ref{tab:cider_jensen_scf} gives the mean absolute error (MAE) and RMSE of these functionals compared to HF and B3LYP, respectively.

Before examining these results, it is worth noting that KS exact exchange (against which CIDER is trained) and HF exact exchange (against which CIDER is tested for SCF calculations) are different quantities yielding different effective potentials~\cite{Gorling1995}. To illustrate why this is, consider that in mean-field theory, the Hamiltonian matrix elements $\mel{\mu}{\hat{H}}{\nu}$ must be evaluated for some basis set $\{\chi_{\mu}(\mathbf{r})\}$ (assumed to be real for simplicity). The KS and HF exchange potential matrix elements are, respectively,
\begin{align}
    \mel{\mu}{\hat{v}_x^{KS}}{\nu}&= \int \diff^3\mathbf{r}\, \chi_{\mu}(\mathbf{r})\chi_{\nu}(\mathbf{r})\fdv{E_x[n]}{n(\mathbf{r})}\\
    \mel{\mu}{\hat{v}_x^{HF}}{\nu}&= -\frac{1}{2} \int \diff^3\mathbf{r}\, \diff^3\mathbf{r}'\, \chi_{\mu}(\mathbf{r})\chi_{\nu}(\mathbf{r}') \frac{n_1(\mathbf{r},\mathbf{r}')}{|\mathbf{r}-\mathbf{r}'|}
\end{align}
These two forms of the exchange matrix elements are distinct. Therefore, the HF and KS exchange energies, potentials, and densities are  different, though this difference tends to be small (about 0.03-0.04\% of the total exchange energy for isolated atoms)~\cite{Gorling1995}. Because exact computation of $\fdv{E_x[n]}{n(\mathbf{r})}$ is complicated and computationally expensive~\cite{Kummel2008}, most modern hybrid DFT calculations (including those performed in this work) use the Generalized Kohn-Sham (GKS) scheme~\cite{Seidl1996,Gorling1997}, in which the HF matrix elements are used for the exact exchange potential instead of the pure KS matrix elements. Because the difference between HF and KS exchange is small, and because evaluating $\fdv{E_x[n]}{n(\mathbf{r})}$ is difficult, it is reasonable to compare CIDER exchange to HF exchange. A more detailed explanation of exchange functionals in the GKS scheme is provided in Supporting Information Section S5 using the Levy-Lieb constrained search formalism~\cite{Levy1979,Levy1982,Lieb1983}.

As shown in Table \ref{tab:cider_jensen_scf}, CIDER-X-AHW has an RMSE of 12 kcal/mol, which is fully explained by the RMSE of the CIDER-X-AHW predictions for static densities of 14 kcal/mol (Figure \ref{fig:cider_jensen_static}). It is notable that the SCF atomization energy errors slightly improve on the atomization exchange energy errors for static densities, even though the functional was only trained on static densities. This suggests that the CIDER exchange potential is sensible.

The 14 kcal/mol error of the CIDER-X-AHW functional with respect to exact exchange is small compared to that of the semi-local exchange functionals investigated here (Figure \ref{fig:cider_jensen_static}), but large compared to the desired chemical accuracy of 1 kcal/mol. However, hybrid functionals often use a small fraction of exact exchange, e.g. 20\% for B3LYP. Because of this, B3LYP-CIDER reproduces B3LYP with an RMSE of only 2.3 kcal/mol on the test set (Table \ref{tab:cider_jensen_scf}). While greater than 1 kcal/mol, this error is promisingly small considering that many of the atomization energies in the Jensen database involve breaking several chemical bonds at once.

\begin{figure}
    \centering
    \includegraphics[width=0.8\textwidth]{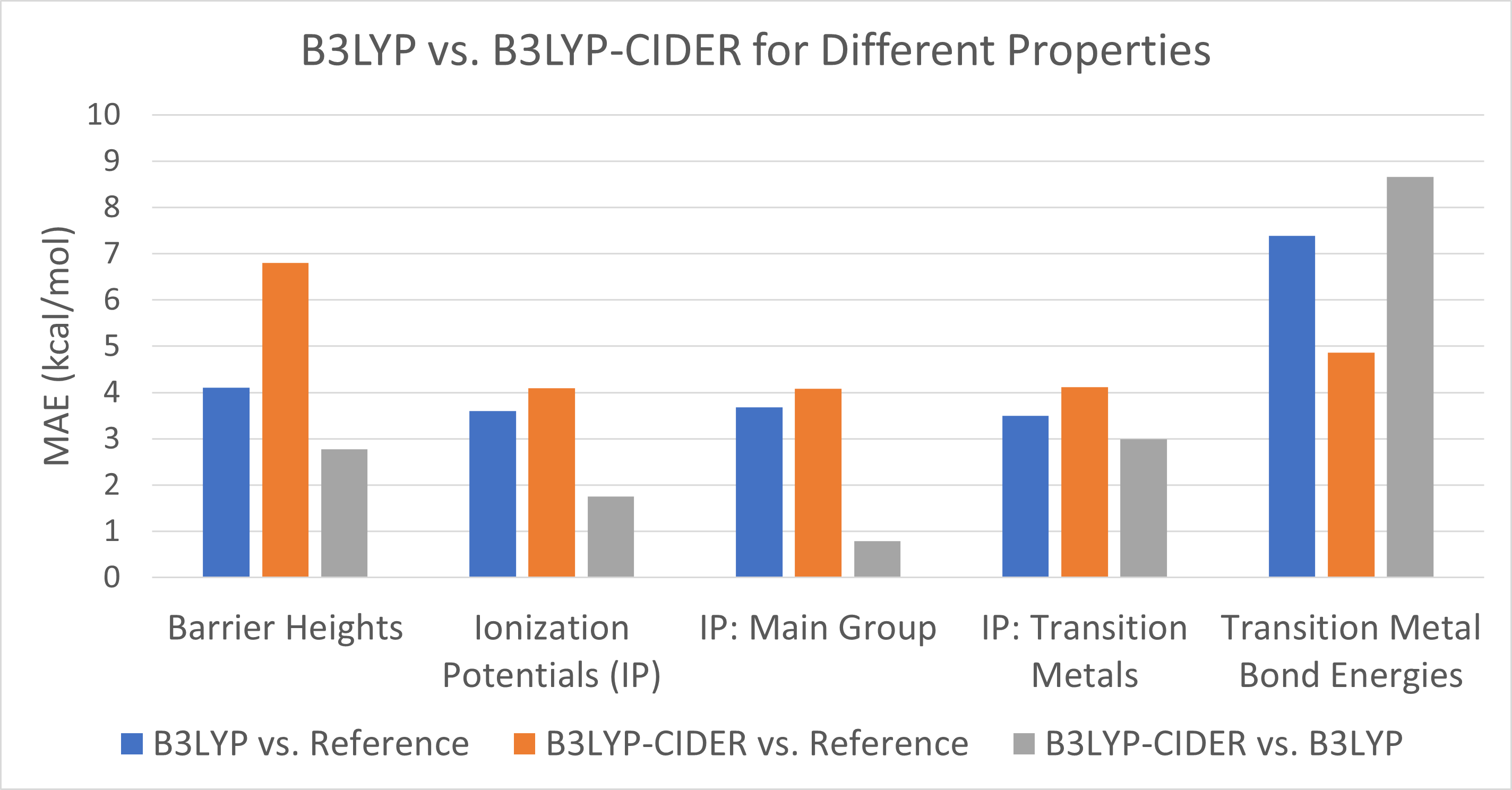}
    \caption{MAE (kcal/mol) of the B3LYP and B3LYP-CIDER functionals compared to both each other and the reference values~\cite{Yu2016MN15} for the BH76 (barrier heights), IP23 (ionization potentials), and TMBE33 (transition metal bond energies) databases.}
    \label{fig:cider_b3lyp_props}
\end{figure}

To test the accuracy of the CIDER functional for both properties and systems significantly different than those contained in the training set, B3LYP and B3LYP-CIDER were used to compute the barriers heights of the BH76 database, the ionization potentials of the IP23 database, and the transition metal bond energies of the TMBE33 database. These databases are subsets of the Minnesota 2015 Database~\cite{Yu2016MN15}. The MAEs for the transition metal bond energies in Figure \ref{fig:cider_b3lyp_props} are divided by the average number of bonds broken per data point, which matches the presentation in the original work~\cite{Yu2016MN15}.

As shown in Figure \ref{fig:cider_b3lyp_props}, B3LYP-CIDER exacerbates the systematic underestimation of barrier heights of B3LYP, leading to an increase in MAE from 4.1 kcal/mol to 6.8 kcal/mol compared to the reference values. The MAE between B3LYP-CIDER and B3LYP is 2.8 kcal/mol. The deviation from B3LYP could potentially be improved by including the density distributions of transition states in the CIDER training set, as currently all training set systems are isolated atoms or molecules at equilibrium geometry.

For ionization potentials, B3LYP-CIDER performs well; the MAE compared to B3LYP is 1.7 kcal/mol, and the error compared to reference values is worse by only 0.5 kcal/mol. Most of the error compared to B3LYP arises from the transition metal systems. For the main-group IP13-MG dataset, B3LYP-CIDER reproduces B3LYP to chemical accuracy (0.8 kcal/mol), while the functionals deviate by 3.0 kcal/mol for the transition metal IP10-TM subset. The chemically accurate reproduction of main-group IPs is notable because B3LYP-CIDER was not trained on any ionic systems. It might be that the uniform scaling rule allows the functional to relate ionic densities to atomic ones, e.g. relating \ce{C+} to \ce{B} because they have the same shell structure.

For transition metal bond energies, B3LYP-CIDER has an MAE of 8.7 kcal/mol relative to B3LYP. However, B3LYP-CIDER is \emph{more} accurate than B3LYP compared to the reference values; the error of the CIDER functional seems to cancel some systematic error of B3LYP for metal bond energies. It is known to be challenging to describe transition metal bonds with hybrid functionals because the semi-local functional's cancellation of error is lost~\cite{Verma2020}. While ideally the CIDER functional would describe the exchange energy accurately, it is reassuring to see that it behaves like a semi-local functional when its accuracy breaks down, leading to sensible and explainable behavior.

Of note, the IP10-TM and TMBE33 datasets include ions, bonded transition metal systems, and fifth-period elements treated with effective core potentials, none of which are present in the training set for CIDER. While CIDER is clearly a less faithful reproduction of HF for these systems than for systems similar to the training set, the resulting errors compared to reference values are similar. In addition, other than the isolated \ce{Fe} atom and \ce{Pd^+} ion, which had to be treated with level shifting and a high damping factor, calculations using the CIDER exchange functional converged successfully, as shown in Table \ref{tab:convergence}. This suggests that the structure of the CIDER functional encourages transferability and stability.

\begin{table}[ht]
    \centering
    \caption{The threshold to which B3LYP-CIDER SCF calculations converged, in Ha, for the 451 systems involved in this study.}
    \begin{tabular}{cr}
        \hline
        Convergence Thresh. & No. of Systems \\
        \hline
        $10^{-9}$ & 425 \\
        $10^{-8}$ & 18 \\
        $10^{-7}$ & 3 \\
        $10^{-6}$ & 3 \\
        Unconverged$^*$ & 2 \\
        \hline
        Total & 451 \\
        \hline
    \end{tabular}\\
    $^*$The two ``unconverged'' calculations were the \ce{Pd+} ion and \ce{Fe} atom, which would only converge to $10^{-7}$ Ha when level shifting was applied.
    \label{tab:convergence}
\end{table}

Lastly, it is worth noting that the CIDER-X-AHW functional was not trained to any total energies, only exchange energy densities. It is possible that retraining the coefficients of hybrid functionals specifically for use with CIDER could significantly improve their accuracy, especially when the functional includes a higher fraction of exact exchange or already has a systematic error for some properties or systems (like B3LYP for barrier heights). An additional consequence of training to the exchange energy density is that it is nontrivial to extend the methodology presented here to the correlation functional. This is because the correlation energy density would be much harder to compute and more ambiguously defined than the exchange energy density. However, this is a limitation of the current Gaussian process model and training procedure, not of the CIDER features themselves. To train a correlation functional, one could either develop an approach to train to total correlation energies with a Gaussian process or use a model for which it is easier to train to total energies, such a neural network.

\subsection{Comment on Computational Cost}

The cost of evaluating the ML model is insignificant because it is mapped to a cubic spline, and the computational bottleneck is the evaluation of the features. Because this initial CIDER model has a large feature set, uses a relatively dense integration grid, and lacks matrix element screening for scalability, feature evaluation is slow for practical applications. However, all of the nonlocal features used in the model are orbital-independent and have a finite length-scale, suggesting that linear scaling and efficient implementations are possible. The challenge of optimizing CIDER functionals to have near-semi-local DFT cost will be the subject of future work. As a first step toward improving the computational efficiency, the Supporting Information (Section S3) introduces a prospective algorithm for evaluating CIDER functionals in a linear-scaling fashion. This algorithm will be implemented in a future work.

%% file: body/conclusion.tex
\section{Conclusion\label{sec:conclusion}}

In this work, we presented the CIDER formalism, which is based on a set of nonlocal features to describe the density distribution in a scale-invariant manner. This feature set was used to train a Gaussian process regression model to accurately describe the Kohn-Sham exchange functional $E_x[n]$, to a level of precision previously only attained by exact evaluation of the functional. The CIDER functional can replace a small fraction of HF exchange in hybrid functionals to accurately reproduce atomization energies, and it has excellent numerical stability, which has previously been a challenge for ML functionals.

As it stands, CIDER demonstrates that smooth, numerically stable exchange functionals can be learned that satisfy known exact constraints, accurately reproduce the target energy, and can be applied across a broad range of the periodic table. This demonstration is a first step towards a functional that could help bridge the gap between efficient semi-local functionals and accurate hybrid functionals, as well as provide a groundwork for developing XC functionals with post-hybrid DFT accuracy.

%% file: body/appendix_extra_math.tex
\section{Relationship Between CIDER Length-Scale and Slater Orbital Density\label{app:math_slater}}

Consider eq \ref{eq:anl} in the case that $B_0=C_0=\frac{6}{5\pi}(6\pi^2)^{2/3}$. Then, noting that $\tau_0=\frac{3}{10}(3\pi^2)^{2/3}n^{5/3}$, eq \ref{eq:anl} becomes
\begin{align}
    a[n](\mathbf{r}) &= \pi \left(\frac{n}{2}\right)^{2/3} C_0 \frac{\tau}{\tau_0}\\
    &= \frac{4\tau}{n}
\end{align}
Now suppose that the density distribution is a spin-unpolarized, two-electron system, in which case $\tau=\frac{|\nabla n|^2}{8n}$. Then
\begin{equation}
    a[n](\mathbf{r})=\frac{1}{2}\left(\frac{|\nabla n|}{n}\right)^2
\end{equation}
If the electron pair occupies a Slater-type orbital, then
\begin{equation}
    n(\mathbf{r})=\frac{\sigma^3}{4\pi}\mathrm{e}^{-\sigma r}\label{eq:sto_dens}
\end{equation}
for some $\sigma$. This orbital has $|\nabla n|/n=\sigma$, so
\begin{equation}
    a[n](\mathbf{r})=\frac{\sigma^2}{2}\label{eq:asigma_rel}
\end{equation}
Then it holds from eqs \ref{eq:sto_dens} and \ref{eq:asigma_rel} that $n(\mathbf{r})\propto \mathrm{e}^{-(\sqrt{2a})r}$, as asserted in eq \ref{eq:slater_orb}.

%% file: body/appendix_extra_math2.tex
\section{Contracting Two $l=1$ Features and One $l=2$ Feature into an $l=0$ Feature\label{app:math_cg}}

Take tensor features $\mathbf{a}, \mathbf{b}, \mathbf{c}$, with $l=1,2,1$, respectively, defined with real spherical harmonics ($x,y,z$ for $l=1$ and $xy,yz,z^2,xz,x^2-y^2$ for $l=2$). We can convert these features to and from the space of complex spherical harmonics using the following rules, with $Y_{lm}$ the real spherical harmonics and $Y_{l}^{m}$ the complex ones:
\begin{align}
    Y_{lm}=\begin{cases}
    \frac{i}{\sqrt{2}}(Y_l^m-(-1)^mY_l^{-m}) & m < 0\\
    Y_l^0 & m = 0\\
    \frac{1}{\sqrt{2}}(Y_l^{-m}+(-1)^mY_l^m) & m > 0
    \end{cases}\\
    Y_l^m=\begin{cases}
    \frac{1}{\sqrt{2}}(Y_{l|m|}-iY_{l,-|m|}) & m < 0\\
    Y_{l0} & m = 0\\
    \frac{(-1)^m}{\sqrt{2}}(Y_{l|m|}+iY_{l,-|m|}) & m > 0
    \end{cases}
\end{align}
This allows us to contract $\mathbf{b}$ and $\mathbf{c}$ to an $l=1$ feature in the complex feature space.
\begin{align}
    d_{-1}&=\sqrt{\frac{6}{10}} b_{-2} c_{+1} - \sqrt{\frac{3}{10}} b_{-1} c_0 + \sqrt{\frac{1}{10}} b_0 c_{-1}\\
    d_0 &= \sqrt{\frac{3}{10}} b_{-1} c_{+1} - \sqrt{\frac{4}{10}} b_{0} c_0 + \sqrt{\frac{3}{10}} b_{+1} c_{-1}\\
    d_{+1} &= \sqrt{\frac{6}{10}} b_{+2} c_{-1} - \sqrt{\frac{3}{10}} b_{+1} c_0 + \sqrt{\frac{1}{10}} b_0 c_{+1}
\end{align}
Then, in eq \ref{eq:feature_vector}, $C(\mathbf{a},\mathbf{b},\mathbf{c})=a_xd_x+a_yd_y+a_zd_z$.

%% file: body/appendix_mapping.tex
\section{Proof That Additive Kernels Have Mappable Predictive Means\label{app:spline_map}}

Consider a kernel with the following structure:
\begin{equation}
    k(\mathbf{x},\mathbf{x}')=\sum_{\{i\}_n} \prod_{d=1}^{n} k_{i_d}(x_{i_d},x_{i_d}')
\end{equation}
This is a general sum of kernels of order $n$, with combinations of $n$ features $\{i\}_n$ used as input to the kernels. For $M$ training points, the predictive mean is determined by the learned weights $\alpha_j$:
\begin{equation}
    f(\mathbf{x}')=\sum_j\alpha_j k(\mathbf{x}',\mathbf{x}^{(j)}).
\end{equation}
By expanding $k(\mathbf{x}',\mathbf{x}^{(j)})$ and switching the order of summations, it can be seen that $f(\mathbf{x}')$ is a sum of functions of the subsets of descriptors:
\begin{align}
    f(\mathbf{x}') &= \sum_j \alpha_j \sum_{\{i\}_n} \prod_{d=1}^{n} k_{i_d}(x_{i_d}',x_{i_d}^{(j)})\\
    f(\mathbf{x}') &= \sum_{\{i\}_n} \left( \sum_j \alpha_j \prod_{d=1}^{n} k_{i_d}(x_{i_d}',x_{i_d}^{(j)})\right)\\
    f(\mathbf{x}') &= \sum_{\{i\}_n} g_{\{i\}_n}\left(x_{\{i\}_n}'\right)\\
    g_{\{i\}_n}\left(x_{\{i\}_n}'\right) &=  \sum_j \alpha_j \prod_{d=1}^{n} k_{i_d}(x_{i_d}',x_{i_d}^{(j)})
\end{align}
Therefore, the predictive mean function can be mapped to a sum of functions of dimension $n$. If $n\le 4$, these component functions can be interpolated using cubic splines, which makes derivative evaluation easy and brings the computational cost of evaluation down to $O(1)$ per test point.

%% file: body/appendix_derivs.tex
\section{Functional Derivatives for CIDER Nonlocal Features\label{app:derivs}}

The Gaussian process is a function of a set of features
\begin{equation}
    e_x[n](\mathbf{r}) = -\frac{3}{4} \left(\frac{3}{\pi}\right)^{1/3} (n(\mathbf{r}))^{4/3} F_x(\mathbf{x}[n](\mathbf{r}))
\end{equation}
with the total exchange energy being
\begin{equation}
    E_x[n] = \int \diff^3\mathbf{r}\, e_x[n](\mathbf{r})
\end{equation}
For the remainder of this section, we denote terms like $\mathbf{x}[n]$ as $\mathbf{x}$ for brevity. Calculating the exchange potential requires functional derivatives with respect to $n(\mathbf{r})$, $\partial_{\alpha} n(\mathbf{r})$ for $\alpha=x,y,z$, and $\tau(\mathbf{r})$. Existing routines in \textsc{PySCF}~\cite{Sun2018} and other DFT codes can compute the Generalized Kohn-Sham potential from the functional derivatives with respect to these quantities.

These functional derivatives can be written as
\begin{align}
    \fdv{E_x}{n(\mathbf{r})} =& - \left(\frac{3}{\pi}\right)^{1/3} 
    n(\mathbf{r})^{1/3} F_x(\mathbf{x}(\mathbf{r}))\notag\\
     &- \frac{3}{4} \left(\frac{3}{\pi}\right)^{1/3} \int \diff^3\mathbf{r}'\, n(\mathbf{r}')^{4/3}
    \left( \sum_i \pdv{F_x}{x_i}\biggr\rvert_{\mathbf{x}(\mathbf{r}')}
    \fdv{x_i}{{n(\mathbf{r})}}\biggr\rvert_{\mathbf{r}'} \right)\label{eq:cider_deriv_rho}\\
    \fdv{E_x}{(\partial_{\alpha}n(\mathbf{r}))} =& - \frac{3}{4} \left(\frac{3}{\pi}\right)^{1/3} \int \diff^3\mathbf{r}'\, n(\mathbf{r}')^{4/3}
    \left( \sum_i \pdv{F_x}{x_i}\biggr\rvert_{\mathbf{x}(\mathbf{r}')}
    \fdv{x_i}{(\partial_{\alpha} n(\mathbf{r}))}\biggr\rvert_{\mathbf{r}'} \right)\label{eq:cider_deriv_grad}\\
    \fdv{E_x}{\tau(\mathbf{r})} =& - \frac{3}{4} \left(\frac{3}{\pi}\right)^{1/3} \int \diff^3\mathbf{r}'\, n(\mathbf{r}')^{4/3}
    \left( \sum_i \pdv{F_x}{x_i}\biggr\rvert_{\mathbf{x}(\mathbf{r}')}
    \fdv{x_i}{{\tau(\mathbf{r})}}\biggr\rvert_{\mathbf{r}'} \right)\label{eq:cider_deriv_tau}
\end{align}
The terms $\pdv{F_x}{x_i}$ are provided by the Gaussian process or cubic spline. If the index $i$ corresponds to a semi-local descriptor, then
\begin{equation}
    \fdv{x_i}{{n(\mathbf{r})}}\biggr\rvert_{\mathbf{r}'}=\pdv{x_i}{n}\biggr\rvert_{n(\mathbf{r})}\delta(\mathbf{r}-\mathbf{r}')\label{eq:slderiv}
\end{equation}
and the integral over $\mathbf{r}'$ reduces to evaluating the derivatives at $\mathbf{r}$.

For the CIDER model, the $G_{nlm}$ descriptors (eq \ref{eq:gnl}) only have nonlocal dependence on the density $n(\mathbf{r}
)$; the dependence on $\tau(\mathbf{r})$ is local, and $G_{nlm}$ does not depend on $\partial_{\alpha} n(\mathbf{r})$. Therefore, eq \ref{eq:slderiv} applies with $\tau(\mathbf{r})$ in place of $n(\mathbf{r})$, even if $x_i$ is nonlocal. The kinetic term $\pdv{x_i}{\tau}\bigr\rvert_{\tau(\mathbf{r})}=\pdv{G_{nlm}}{\tau}\bigr\rvert_{\tau(\mathbf{r})}$, needed to evaluate eq \ref{eq:cider_deriv_tau}, arises solely from the derivative of the exponent:
\begin{align}
    \pdv{G_{nlm}}{\tau}\biggr\rvert_{\tau(\mathbf{r})} =& \pdv{G_{nlm}}{a}\biggr\rvert_{a(\mathbf{r})} \pdv{a}{\tau}\biggr\rvert_{\tau(\mathbf{r})}\\
    \pdv{G_{nlm}}{a} =& \frac{l}{2a}G_{nlm}(\mathbf{r})
    - H_{nlm}(\mathbf{r}) \\
    \pdv{a}{\tau} =& C_0\pi\left(\frac{n}{2}\right)^{2/3}\frac{1}{\tau_0}\\
    H_{nlm}(\mathbf{r}) =& \int \diff^3\mathbf{r}'\,\, |\mathbf{r}'-\mathbf{r}|^2\,g_{nlm}(\mathbf{r}'-\mathbf{r}; \mathbf{r})\, n(\mathbf{r}')\label{eq:hnl}
\end{align}
The density derivatives are similar, but with an additional nonlocal term $g_{nlm}(\mathbf{r}-\mathbf{r}';\mathbf{r}')$:
\begin{align}
    \fdv{G_{nlm}}{n(\mathbf{r})}\biggr\rvert_{\mathbf{r}'} =& \pdv{G_{nlm}}{n}\biggr\rvert_{n(\mathbf{r})}\delta(\mathbf{r}-\mathbf{r}') + g_{nlm}(\mathbf{r}-\mathbf{r}';\mathbf{r}') \label{eq:rho_fderiv} \\
    \pdv{G_{nlm}}{n}\biggr\rvert_{n(\mathbf{r})} =& \pdv{G_{nlm}}{a}\biggr\rvert_{a(\mathbf{r})} \pdv{a}{n}\biggr\rvert_{n(\mathbf{r})}\\
    \pdv{a}{n} =& (C_0-B_0)\pi\left(\frac{1}{4n}\right)^{1/3}\frac{\tau}{\tau_0}
\end{align}
The second term on the right hand side of eq \ref{eq:rho_fderiv} introduces a nonlocal term $v_{nlm}(\mathbf{r})$ in eq \ref{eq:cider_deriv_rho} of the form
\begin{align}
    v_{nlm}(\mathbf{r}) &= \int \diff^3\mathbf{r}'\, f_{nlm}(\mathbf{r}')
    g_{nlm}(\mathbf{r}-\mathbf{r}';\mathbf{r}') \label{eq:deriv_nonlocal}\\
    f_{nlm}(\mathbf{r}) &= -\frac{3}{4} \left(\frac{3}{\pi}\right)^{1/3} n(\mathbf{r})^{4/3}
    \pdv{F_x}{G_{nlm}}\biggr\rvert_{\mathbf{x}(\mathbf{r})}
\end{align}

To obtain the Generalized Kohn-Sham potential, one must compute the matrix elements for an atomic orbital basis $\{\ket{\mu}\}$ (such that $\braket{\mathbf{r}}{\mu}=\chi_{\mu}(\mathbf{r})$):
\begin{equation}
\begin{split}
    v_{\mu\nu} =& \int \diff^3\mathbf{r}\, \chi_{\mu}(\mathbf{r}) \fdv{E_x}{n(\mathbf{r})} \chi_{\nu}(\mathbf{r})\\
    &+ \sum_{\alpha=x,y,z} \int \diff^3\mathbf{r}\, \chi_{\mu}(\mathbf{r}) \fdv{E_x}{(\partial_{\alpha} n(\mathbf{r}))} \partial_{\alpha}\chi_{\nu}(\mathbf{r})\\
    &+ \sum_{\alpha=x,y,z} \int \diff^3\mathbf{r}\, \chi_{\nu}(\mathbf{r}) \fdv{E_x}{(\partial_{\alpha} n(\mathbf{r}))} \partial_{\alpha}\chi_{\mu}(\mathbf{r})\\
    &+ \frac{1}{2}\int \diff^3\mathbf{r}\, \left(\nabla\chi_{\mu}(\mathbf{r})\cdot \nabla \chi_{\nu}(\mathbf{r}) \right) \fdv{E_x}{\tau(\mathbf{r})}
\end{split}
\label{eq:ksmel}
\end{equation}
The above equation assumes real orbitals. Typically, the integrals over $\mathbf{r}$ are evaluated numerically on a real-space grid. Because their contributions are local, all terms except for eq \ref{eq:deriv_nonlocal} can be evaluated with this standard numerical integration approach. Equation \ref{eq:deriv_nonlocal} gives rise to a double integration in eq \ref{eq:ksmel}:
\begin{align}
    v_{\mu\nu}^{nlm} = &\int \diff^3\mathbf{r}\, \left[\chi_{\mu}(\mathbf{r})\chi_{\nu}(\mathbf{r})\right] \notag\\
    &\times \int \diff^3\mathbf{r}'\, f_{nlm}(\mathbf{r}')
    g_{nlm}(\mathbf{r}-\mathbf{r}';\mathbf{r}')\label{eq:nlderiv}
\end{align}

This analysis leaves three nonlocal terms that must be evaluated at each iteration: $G_{nlm}(\mathbf{r})$ (eq \ref{eq:gnl}), $H_{nlm}(\mathbf{r})$ (eq \ref{eq:hnl}), and $v_{\mu\nu}^{nlm}$ (eq \ref{eq:nlderiv}). This is done using a density fitting (DF) auxiliary basis $\{\Theta_p(\mathbf{r})\}$,
\begin{equation}
    \chi_{\mu}(\mathbf{r})\chi_{\nu}(\mathbf{r})=\sum_p C_p^{\mu\nu} \Theta_p(\mathbf{r})
\end{equation}
Then the nonlocal terms can be evaluated in a computationally efficient manner,
\begin{align}
    G_{nlm}(\mathbf{r}) &= \sum_p n_p \braket{g_{nlm}(\mathbf{r})}{\Theta_p}\\
    H_{nlm}(\mathbf{r}) &= \sum_p n_p \mel{g_{nlm}(\mathbf{r})}{|\mathbf{r}'-\mathbf{r}|^2}{\Theta_p}\\
    v_{\mu\nu}^{nlm} &= \sum_p C_p^{\mu\nu} v_p^{nlm}\\
    v_p^{nlm} &= \int \diff^3\mathbf{r} f_{nlm}(\mathbf{r}) \braket{g_{nlm}(\mathbf{r})}{\Theta_p}
\end{align}
The integral over $\mathbf{r}$ in the last equation is evaluated numerically; all braket notation terms are evaluated analytically. The density $n_p$ in the auxiliary basis is computed from the atomic orbital density matrix $P_{\mu\nu}$, which in turn is evaluated from the KS orbitals $\phi_i$ and occupations $f_i$:
\begin{align}
    n_p &= \sum_{\mu\nu} C_p^{\mu\nu} P_{\mu\nu}\\
    P_{\mu\nu} &= \sum_i f_i c_{i\mu} c_{i\nu}\\
    \phi_i(\mathbf{r}) &= \sum_{\mu} c_{i\mu} \chi_{\mu}(\mathbf{r})
\end{align}
The above equations assume real coefficients $c_{i\mu}$.